\documentclass[fleqn,usenatbib]{mnras}

\usepackage[OT2,T1]{fontenc}
\usepackage{ae,aecompl}

\usepackage{times}
\usepackage{epsfig}
\usepackage[]{graphicx} 
\usepackage{pdfpages}
\usepackage{verbatim} 
\usepackage[fleqn]{amsmath}
\usepackage{amssymb}
\usepackage{calc}
\usepackage[normal]{threeparttable}
\usepackage{longtable}
\usepackage{pdflscape}
\newcommand{\comments}[1]{} 
\usepackage{placeins} 
\usepackage{float}
\usepackage{soul}
\usepackage{relsize}
\graphicspath{{./}}
\title[Cold gas towards two lensed quasars]{Illuminating the past
  8\,billion years of cold gas towards two gravitationally lensed
  quasars}

\author[J.~R. Allison et al.]{
J.~R. Allison,$^{1}$\thanks{E-mail: james.allison@csiro.au}
V.~A. Moss,$^{2,3}$
J.-P. Macquart,$^{3,4}$
S.~J. Curran,$^{5}$
S.~W. Duchesne,$^{5}$
\newauthor 
E.~K. Mahony,$^{2,3}$
E.~M. Sadler,$^{2,3}$
M.~T. Whiting,$^{1}$ 
K.~W. Bannister,$^{1}$ 
A.~P. Chippendale,$^{1}$
\newauthor
P.~G. Edwards,$^{1}$
L. Harvey-Smith,$^{1}$
I. Heywood,$^{1,6}$
B.~T. Indermuehle,$^{1}$
E. Lenc,$^{1,2,3}$
\newauthor
J. Marvil,$^{1}$
D. McConnell,$^{1}$
and
R.~J. Sault$^{1,7}$
\\
$^{1}$CSIRO Astronomy and Space Science, PO Box 76, Epping NSW 1710,
Australia\\
$^{2}$Sydney Institute for Astronomy, School of Physics A28, University of
Sydney, NSW 2006, Australia\\
$^{3}$ARC Centre of Excellence for All-sky Astrophysics (CAASTRO)\\
$^{4}$ICRAR/Curtin University, Curtin Institute of Radio Astronomy, Perth, WA
6845, Australia\\
$^{5}$School of Chemical and Physical Sciences, Victoria University of
Wellington, PO Box 600, Wellington 6140, New Zealand\\
$^{6}$Department of Physics and Electronics, Rhodes University, PO
Box 94, Grahamstown 6140, South Africa \\
$^{7}$School of Physics, University of Melbourne, VIC 3010,
Australia}

\date{Accepted XXX. Received YYY; in original form ZZZ}

\pubyear{2016}

\begin{document}
\label{firstpage}
\pagerange{\pageref{firstpage}--\pageref{lastpage}}
\maketitle

\begin{abstract}
Using the Boolardy Engineering Test Array of the Australian Square
  Kilometre Array Pathfinder (ASKAP BETA), we have carried out the
  first $z = 0 - 1$ survey for \mbox{H\,{\sc i}} and OH absorption
  towards the gravitationally lensed quasars PKS\,B1830$-$211 and
  MG\,J0414$+$0534. Although we detected all previously reported intervening
  systems towards PKS\,B1830$-$211, in the case of MG\,J0414+0534 three
  systems were not found, indicating that the original identifications
  may have been confused with radio frequency interference. Given the
  sensitivity of our data, we find that our detection yield is
  consistent with the expected frequency of intervening \mbox{H\,{\sc
  i}} systems estimated from previous surveys for 21-cm emission in
  nearby galaxies and $z \sim 3$ damped Lyman $\alpha$ absorbers. We
  find spectral variability in the $z = 0.886$ face-on spiral galaxy
  towards PKS\,B1830$-$211, from observations undertaken with the
  Westerbork Synthesis Radio Telescope in 1997/1998 and ASKAP BETA in
  2014/2015. The \mbox{H\,{\sc i}} equivalent width varies by a few
  per cent over approximately yearly time-scales. This long-term
  spectral variability is correlated between the north-east and south-west images of the core, and with the total flux density of the
  source, implying that it is observationally coupled to intrinsic
  changes in the quasar. The absence of any detectable
  variability in the ratio of \mbox{H\,{\sc i}} associated with the two core images is in stark contrast to the behaviour
  previously seen in the molecular lines. We therefore infer that
  coherent opaque \mbox{H\,{\sc i}} structures in this galaxy are
  larger than the parsec-scale molecular clouds found at
  mm-wavelengths.
\end{abstract}

\begin{keywords} 
  galaxies: evolution -- galaxies: high redshift -- galaxies: ISM -- quasars: absorption lines -- galaxies: structure -- radio lines:
  galaxies
\end{keywords}



\section{Introduction}\label{intro}

The neutral interstellar medium (ISM) plays a fundamental role in galaxy
evolution, providing the fuel available for star formation
(e.g. \citealt{Schmidt:1959, Kennicutt:1998}) and
the growth of supermassive black holes (see
\citealt{Heckman:2014} for a review). Determining how the physical
state of the neutral ISM in galaxies changes over
cosmological time-scales is therefore critical to our understanding of
how these processes evolve throughout the history of the Universe. Neutral atomic hydrogen (\mbox{H\,{\sc i}}) is typically abundant in gas-rich galaxies and readily detectable
via either the 21-cm line in the nearby Universe or the Lyman $\alpha$
line at high redshifts. \mbox{H\,{\sc i}} is therefore often used as a tracer of the neutral gas throughout the history of the Universe
(see \citealt{Wolfe:2005, Giovanelli:2016} for
reviews). However, the evolution of the \mbox{H\,{\sc i}} mass density over cosmological time scales (e.g. \citealt{Zwaan:2005,
  Martin:2010, Braun:2012, Noterdaeme:2012,  Zafar:2013, Crighton:2015,
  Sanchez-Ramirez:2016}) appears to be weaker by an order of magnitude than either the total star
formation rate (e.g. \citealt{Hopkins:2006, Sobral:2013,
  Gunawardhana:2013, Burgarella:2013, Zwart:2014}) or the molecular
gas (e.g. \citealt{Keres:2003, Carilli:2013}). It is therefore likely that
the \mbox{H\,{\sc i}} in galaxies represents a quasi-steady intermediate state
between
the warm ionized and cold molecular gas, with the latter 
responsible for directly fuelling star formation
(e.g. \citealt{Lagos:2014}). Although the 
\mbox{H\,{\sc i}} may not directly trace the formation of stars
in galaxies, understanding how its physical state, in particular the
cold neutral medium (CNM; $\sim$100\,K), evolves over cosmological time will
help elucidate its role in the formation of molecular gas and future star formation.

The present census of \mbox{H\,{\sc i}} at intermediate cosmological redshifts ($z \approx 0.2 - 1.7$) is limited by significant observational challenges. The 21-cm line emission from individual galaxies is observationally expensive to detect with existing radio telescopes (see e.g. \citealt{Catinella:2015, Fernandez:2016}) and cosmological work has focused on statistical detection through either stacking (e.g. \citealt{Lah:2007, Kanekar:2016, Rhee:2016}) or intensity mapping (e.g. \citealt{Chang:2010, Masui:2013}). Likewise, the Lyman $\alpha$ line is only observable at ultra-violet (UV) wavelengths and necessitates the use of space-borne telescopes. Using the \emph{Hubble Space Telescope (HST)}, \cite{Rao:2006} carried out a survey of damped Lyman $\alpha$ absorbers (DLAs; $N_{\rm HI} \geq 2 \times 10^{20}\,\mathrm{cm}^{-2}$) through targeted  observations of quasars with known strong \mbox{Mg\,{\sc ii}} absorption. They found that the DLA mass density showed little evolution between $z \approx 5$ and 0.5, but then must decrease by a factor of $\approx$2 in the past 5\,Gyr to be consistent with results from 21-cm surveys. A more recent survey of the \emph{HST} data by \cite{Neeleman:2016} found a significantly lower incidence of DLAs when they accounted for the pre-selection biases due to targeting quasars with known metal absorbers and intervening galaxies. However, in both cases, fractional uncertainties are large due to the relatively small samples sizes involved, compared with nearby 21-cm surveys and larger optical DLA surveys at higher redshifts. We can instead use 21-cm absorption
against suitably bright background radio sources to determine the
abundance and kinematics of the \mbox{H\,{\sc i}}. Furthermore, since 21-cm absorption is particularly sensitive to the cold gas, it is a powerful
additional tool in surveying this phase of the \mbox{H\,{\sc i}} at
high redshift (\citealt{Kanekar:2004, Morganti:2015}).

In this paper, we present results from a 21-cm absorption survey for
cold \mbox{H\,{\sc i}} and OH gas towards two gravitationally lensed
quasars -- PKS\,B1830$-$211 at $z = 2.51$ (\citealt{Lidman:1999}) and
MG\,J0414$+$0534 at $z = 2.64$ (\citealt{Lawrence:1995a}) -- covering
intermediate cosmological redshifts between $z \approx 0$ and 1. We
use commissioning data from the Boolardy Engineering Test Array (BETA;
\citealt{Hotan:2014, McConnell:2016}) of the Australian Square Kilometre Array
Pathfinder (ASKAP; \citealt{Johnston:2007, Schinckel:2012}). The
relatively clean radio frequency environment and large fractional
bandwidth available with this interferometer is particularly
well suited to such a survey. Our choice of targets is based on the
unique nature of the alignment between the radio source and at least
one foreground galaxy. In both cases, the background quasar is strongly
lensed into several compact image components, thereby creating
multiple sight lines through each intervening system and so providing
additional spatial information. Furthermore, the effect of any
temporal behaviour in either the foreground galaxy or the background
quasar will be magnified by the lensing system and is
potentially detectable in the spectral line profile. Evidence for
variability in extragalactic 21-cm absorption lines is rare and has only previously
been found in the spectra of a few quasars (e.g. \citealt{Wolfe:1982,
  Kanekar:2001a}). These examples are attributed to either transverse
motion of jet-knots in the background quasar (e.g \citealt{Briggs:1983}) or interstellar scintillation (ISS)
of a sufficiently compact background source or absorbing structure (e.g. \citealt{Macquart:2005}). The time-scales for spectral variability can then be used to infer the physical scale for optical depth variations 
in the absorbing \mbox{H\,{\sc i}} gas.

\mbox{H\,{\sc i}} absorption has been reported
in the dominant lensing galaxy of each quasar
(\citealt{Chengalur:1999, Curran:2007b}), as well as several other possible
intervening galaxies along the same sight line
(e.g. \citealt{Lovell:1996, Curran:2011c, Tanna:2013}). In the case of
MG\,J0414+0534, the number of intervening systems reported in the literature is far more than expected given the number density of DLAs estimated from large optical surveys (\citealt{Tanna:2013}). However, flux ratios observed between the
brightest image components at radio and mid-infrared wavelengths is
inconsistent with a lens model based on just the detected optical
counterparts, requiring a more complex mass distribution in the form
of either satellite or intervening haloes and filaments
(e.g. \citealt{MacLeod:2013, Inoue:2015}). Therefore, by confirming whether
there are indeed  
intervening \mbox{H\,{\sc i}} systems, we hope that our data will help inform
these
alternative lens models.

We structure this paper as follows. In \autoref{section:observations_data}, we
present a brief description of our observations and data analysis. We discuss
our results for the two quasars in \autoref{section:pks1830-211} and
\autoref{section:mgj0414+0534}. In \autoref{section:expected_number}, we
estimate the expected number of intervening systems and discuss whether our
results are consistent. We summarize our conclusions in
\autoref{section:conclusions}. 

\section{Observations and Data}\label{section:observations_data}

\begin{table*} 
   \caption{Summary of observations using the ASKAP BETA
     telescope. Each observation is assigned a unique scheduling block
     identification (SBID) number. $t_{\rm int}$ denotes the on-source
     integration time (averaged over antenna baselines and taking into
     account flagging). $\sigma_{\rm chan}$ is the median rms
     noise\tnote{$a$} per 18.5\,kHz channel as a percentage of the
     continuum (relevant for absorption detection). Note that the
     shortest baseline (AK01 -- AK03) is discarded during data
     processing (see \citealt{Serra:2015}).}\label{table:beta_observations}
 \begin{threeparttable}
  \begin{tabular}{lllllllcc}
    \hline
    \multicolumn{1}{l}{Source name} & \multicolumn{1}{l}{SBID}  &
\multicolumn{1}{l}{Date} & \multicolumn{2}{c}{MJD} &
\multicolumn{1}{c}{Frequency\,band} & \multicolumn{1}{l}{Antennas\tnote{$b$}} &
\multicolumn{1}{c}{$t_\mathrm{int}$} & \multicolumn{1}{c}{$\sigma_{\rm chan}$}
\\
    \multicolumn{1}{c}{} & \multicolumn{1}{c}{} & \multicolumn{1}{c}{} &
\multicolumn{1}{c}{Start} & \multicolumn{1}{c}{End} & \multicolumn{1}{c}{[MHz]}
& \multicolumn{1}{c}{} & \multicolumn{1}{c}{[h]} &
\multicolumn{1}{c}{[per\,cent]} \\   
    \hline
    PKS\,B1830$-$211 & 163, 165 & 2014 July 07 & 56845.447 & 56845.887 & 711.5
-- 1015.5 & 1, 3, 6, 8, 15 & 9.0 & 0.14 \\ 
    & 166, 167 & 2014 July 08 & 56846.429 & 56846.872 & 967.5 -- 1271.5 & 3, 6,
8, 9, 15 & 8.4 & 0.13 \\ 
    & 213 & 2014 July 11 & 56849.419 & 56849.627 & 1223.5 -- 1527.5 & 1, 3, 6,
8, 9, 15 & 4.0 & 0.33 \\
    & 2216  & 2015 July 24 & 57227.411 & 57227.536 & 711.5 -- 1015.5 & 1, 3, 6,
8, 15 & 2.5 & 0.29 \\
    & 2248 & 2015 July 29 & 57232.573 & 57232.807 & 711.5 -- 1015.5 & 1, 3, 6,
8, 15 & 4.8 & 0.22 \\
    & 2328 & 2015 August 08 & 57242.377 & 57242.757 & 967.5 -- 1271.5 & 1, 6,
8, 15 & 7.2  & 0.19 \\
    & 2640 & 2015 September 28 & 57293.483 & 57293.629 & 711.5 -- 1015.5 & 1,
3, 6, 8, 15 & 2.4 & 0.30 \\
    & 2655 & 2015 September 29 & 57294.442 & 57294.567 & 711.5 -- 1015.5 & 1,
3, 6, 8, 15 & 2.5 & 0.28 \\
    & 2716 & 2015 October 13 & 57308.371 & 57308.587 & 711.5 -- 1015.5 & 1, 3,
6, 8, 15 & 3.8 & 0.24 \\
    & 2747 & 2015 October 18 & 57313.194 & 57313.407 & 711.5 -- 1015.5 & 1, 3,
6, 8, 15 & 4.2 & 0.23 \\
    & 2789 & 2015 October 21 & 57316.192 & 57316.405 & 711.5 -- 1015.5 & 1, 3,
6, 8, 15 & 4.3 & 0.23 \\
    & 2898 & 2015 October 24 & 57319.376 & 57319.584 & 711.5 -- 1015.5 & 1, 3,
6, 8, 15 & 3.5 & 0.25 \\
    & 2925, 2927 & 2015 October 26 & 57321.408 & 57321.560 & 711.5 -- 1015.5 &
1, 3, 6, 8, 15 & 2.4 & 0.31 \\
    & 2936 & 2015 October 27 & 57322.303 & 57322.520 & 711.5 -- 1015.5 & 1, 3,
6, 8, 15 & 4.0 & 0.24 \\
    & 2950 & 2015 October 28 & 57323.426 & 57323.551 & 711.5 -- 1015.5 & 1, 3,
6, 8, 15 & 2.0 & 0.34 \\    
    & 2973, 2974 & 2015 October 30 & 57325.288 & 57325.519 & 711.5 -- 1015.5 &
1, 3, 6, 8, 15 & 4.2 & 0.23 \\
    & & & & & & & & \\ 
    MG\,J0414$+$0534 & 218 & 2014 July 12 & 56850.176 & 56850.237 & 711.5 --
1015.5 & 1, 3, 6, 8, 9, 15 & 1.3 & 1.8 \\
    & 1544 & 2015 March 17 & 57098.279 & 57098.430 & 967.5 -- 1271.5 & 1, 3, 8,
9, 15 & 2.5 & 1.4 \\ 
    & 1545 & 2015 March 17 & 57098.433 & 57098.557 & 1223.5 -- 1527.5 & 1, 3,
8, 9, 15 & 2.1 & 3.6 \\ 
    & 1962, 1966 & 2015 June 16 & 57189.099 & 57189.267 & 711.5 -- 1015.5 & 1,
6, 8, 15 & 2.7 & 1.9 \\ 
    & 2225 & 2015 July 25 & 57228.838 & 57229.204 & 711.5 -- 1015.5 & 1, 3, 6,
8, 15 & 6.9 & 0.76 \\ 
    & 2231 & 2015 July 27 & 57230.815 & 57230.998 & 1223.5 -- 1527.5 & 1, 3, 6,
8, 15 & 3.3 & 2.7 \\ 
    & 2336 & 2015 August 10 & 57244.796 & 57245.110 & 967.5 -- 1271.5 & 1, 6,
8, 15 & 5.6 & 1.2 \\ 
    & 2340 & 2015 August 11 & 57245.830 & 57246.019 & 967.5 -- 1271.5 & 1, 6,
8, 15 & 3.2 & 1.5 \\ 
    & 2938 & 2015 October 27 & 57322.628 & 57322.786 & 711.5 -- 1015.5 & 1, 3,
6, 8, 15 & 2.5 & 1.4 \\
    & 2952 & 2015 October 28 & 57323.660 & 57323.944 & 711.5 -- 1015.5 & 1, 3,
6, 8, 15 & 4.9 & 0.98 \\    
    & 2967 & 2015 October 29 & 57324.809 & 57324.942 & 711.5 -- 1015.5 & 1, 3,
6, 8, 15 & 2.2 & 1.8 \\
    & 3013 & 2015 November 03 & 57329.654 & 57329.928 & 711.5 -- 1015.5 & 1, 3,
6, 8, 15 & 4.7 & 1.1 \\
    & 3046 & 2015 November 06 & 57332.624 & 57332.921 & 711.5 -- 1015.5 & 1, 3,
6, 8, 15 & 5.1 & 0.88\\    
    & 3062 & 2015 November 08 & 57334.587 & 57334.794 & 711.5 -- 1015.5 & 1, 3,
6, 8, 15 & 2.9 & 1.2 \\
    \hline
  \end{tabular}
   \begin{tablenotes}
   \item[$a$] {In general, the noise scales as expected with the
       integration time and number of antennas. However, it should be
       noted that during commissioning, the sensitivity of ASKAP BETA did change between observations and was known to decrease significantly towards the higher frequencies (see e.g. \citealt{McConnell:2016}).}
   \item[$b$] {See figure\,2 of \citet{Hotan:2014} for details of the
       ASKAP BETA antenna positions.}
   \end{tablenotes}
 \end{threeparttable}
\end{table*}

\subsection{Observations with ASKAP BETA}

During the period 2014 July to 2015 November, we carried out several
observations of PKS\,B1830$-$211 and MG\,J0414$+$0534 with the ASKAP
BETA telescope, which we summarize in
\autoref{table:beta_observations}. We followed a similar observing
procedure to that described by \cite{Hotan:2014} and
\cite{McConnell:2016}, whereby focal-array primary beams were
electronically formed to maximize the signal-to-noise ratio (S/N)
towards nominal centre positions. Although the telescope can be used to
form up to nine beams simultaneously for wide-field imaging, in this
paper, we focus on a single beam formed towards the target quasar at
the pointing centre. Observations of each quasar were accompanied by
short (5 -- 15\,min) scans of PKS\,B1934$-$638 for initial calibration of the complex gains and setting the flux scale based on the model of
\cite{Reynolds:1994}\footnote{\url{http://www.atnf.csiro.au/observers/memos/}}.

Three frequency bands were used: 711.5 -- 1015.5\,MHz, 967.5 -- 1271.5\,MHz,
and 1223.5 -- 1527.5\,MHz, spanning all \mbox{H\,{\sc i}} redshifts
between $z = 0$ and 1. We spent more observing time on the lowest
frequency band, which covered most of the redshift interval, was least
contaminated by radio frequency interference (RFI), and contained known
absorption lines which we either wanted to confirm or monitor for
variability. The number of available antennas varied between
observations (see Table\,\ref{table:beta_observations}), but in
general the full six-antenna BETA is sensitive to angular scales in
the range 40\,arcmin to 45\,arcsec between 711.5 and
1527.5\,MHz. Therefore, both the target quasars 
were spatially unresolved in all observations and could be
treated as point-like radio sources. The ASKAP BETA fine filterbank
generated 16\,416 spectral channels separated by approximately
18.5\,kHz, equivalent to velocity resolutions in the range 3.6 --
7.8\,km\,s$^{-1}$ spanning the three frequency bands.

\begin{figure*}
\centering
\includegraphics[width=0.9\textwidth]{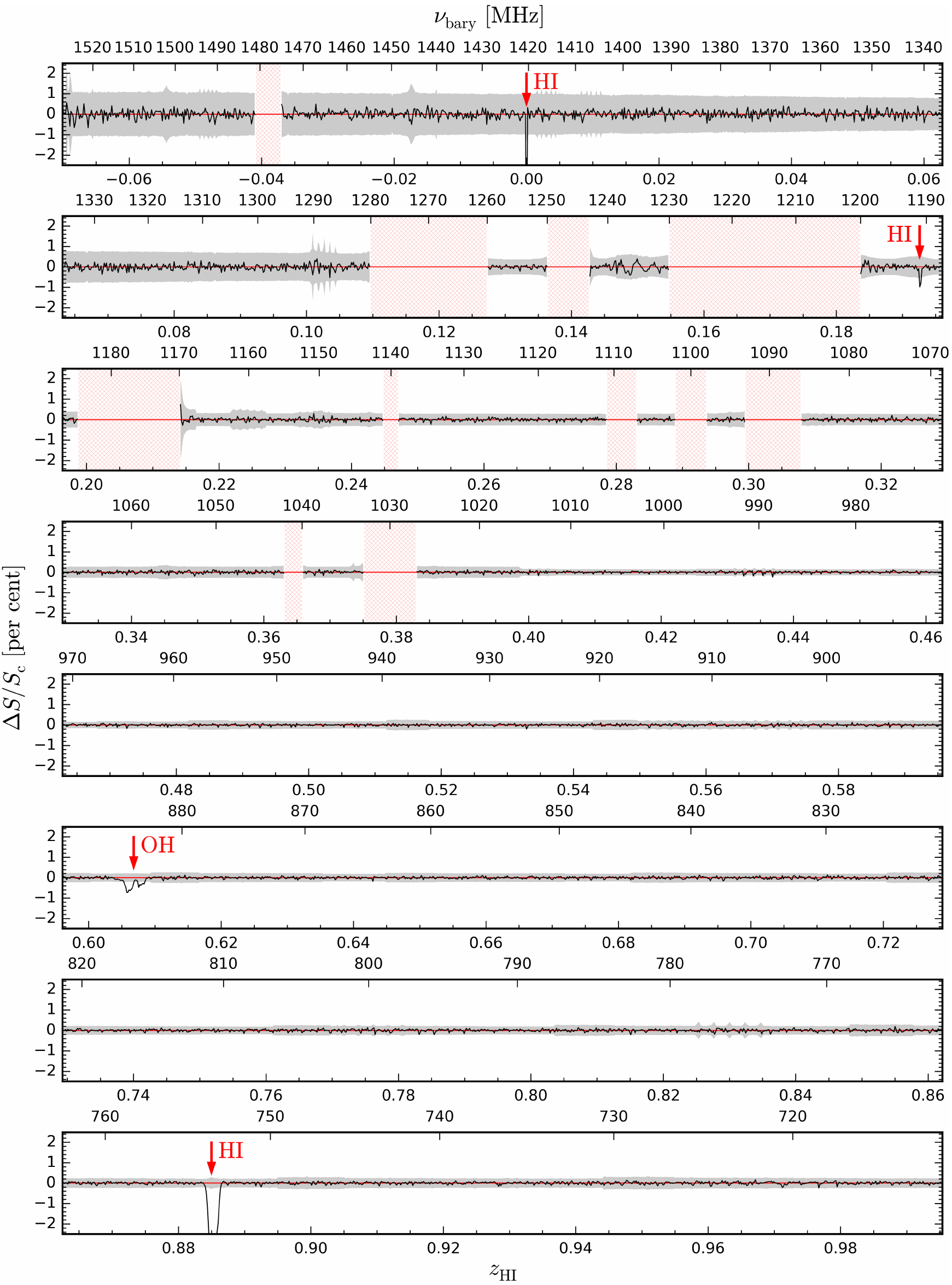}
\caption{The final ASKAP BETA spectrum (averaged over all observations)
  towards PKS\,B1830$-$211. For visual clarity, the data have been
  binned by a factor of 3, from the native spectral resolution of 18.5\,kHz to
  55.6\,kHz. The barycentric \mbox{H\,{\sc i}} redshift is shown on
  the lower horizontal axis and the corresponding observed frequency
  is shown on the upper horizontal axis. The data (black line) denote
  the change in flux density ($\Delta{S}$) as a percentage of the total continuum ($S_{\rm c}$) and the grey region gives the corresponding rms spectral noise multiplied by
  a factor of 5. Prominent features in the noise are the
  result of individual failures in the beamformer or correlator cards for
  specific observations. The hatched regions mask data which were
  significantly contaminated by aviation and satellite-generated RFI. The red arrows indicate the positions of
  previously reported \mbox{H\,{\sc i}} and \mbox{OH}
  lines.}\label{figure:PKS1830-211_beta_spectrum}
\end{figure*}

\begin{figure*}
\centering
\includegraphics[width=0.9\textwidth]{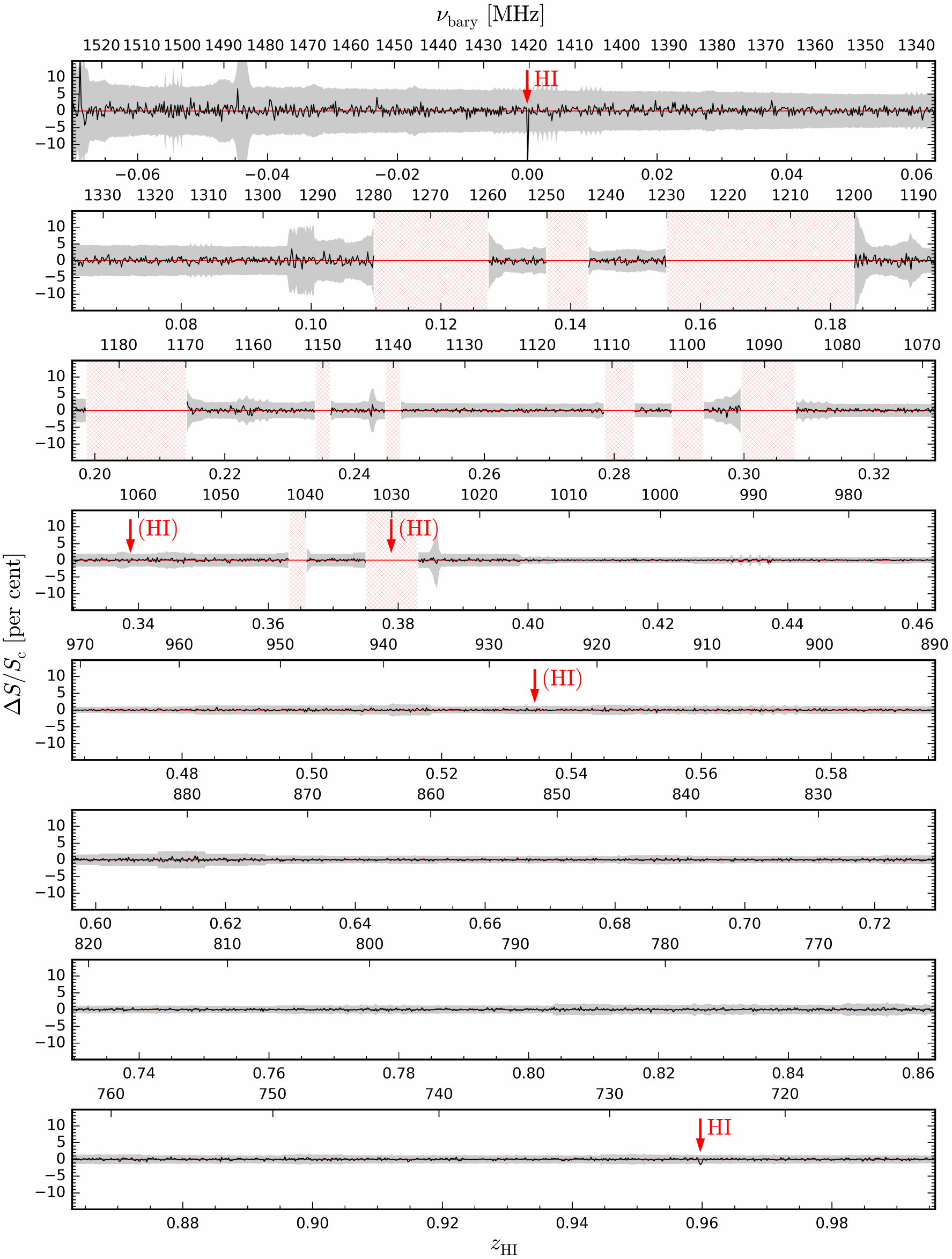}
\caption{As \autoref{figure:PKS1830-211_beta_spectrum}, but showing the final spectrum for MG\,J0414+0534. Parentheses indicate those \mbox{H\,{\sc i}} lines which were previously reported in the literature but were not detected in our data (see text for details).}\label{figure:MGJ0414+0534_beta_spectrum}
\end{figure*}

\subsection{Data reduction}

For calibration, flagging, and imaging of ASKAP BETA data, we followed a procedure similar to that described by \cite{Allison:2015a}, hereafter referred to as A15. The uncalibrated visibilities from the ASKAP correlator were recorded in measurement set format and so we used the \texttt{CASA}\footnote{\url{http://casa.nrao.edu}} package (\citealt{McMullin:2007}) for initial flagging, averaging, and splitting of the data. To aid bandpass calibration, we split the full spectral resolution data into the intervals on which the primary beams were formed (4 and 5\,MHz chunks). This has the additional benefit that the data could then be more efficiently processed using standard data reduction packages in a parallel computing environment. Separately, we produced a frequency-averaged data set spanning the entire 304\,MHz band, which could then used for forming high S/N continuum images for self-calibration. However, to mitigate contamination from satellite- and aviation-generated RFI (seen at the higher frequencies included in this work), we increased the resolution of the averaged data from the 9.5 MHz bins used by A15 to 1 MHz.

All subsequent flagging, calibration and imaging was carried out using standard tasks from the \texttt{MIRIAD}\footnote{\url{http://www.atnf.csiro.au/computing/software/miriad}} package (\citealt{Sault:1995}). Automated flagging of corrupted data was achieved using an implementation of the \texttt{SumThreshold} method (\citealt{Offringa:2010}) in the \texttt{PGFLAG} task, which for broad-band-RFI free regions of the spectrum resulted in data loss of a few percent. Observations of PKS\,B1934-638 were used to set the flux scale, based on the model of \cite{Reynolds:1994}, and solutions for the complex gain were then copied across to the corresponding observation of the quasar. Further corrections to the time varying complex gains were calculated using several iterations of self-calibration on the full-band 1\,MHz averaged data at successively smaller solution intervals. The initial model was constructed from the catalogue of the National Radio Astronomy Observatory Very Large Array Sky Survey (NVSS; \citealt{Condon:1998}), followed by multifrequency \texttt{CLEAN} component models constructed from imaging the data. The gain solutions were then transferred to all 4 and 5\,MHz chunks of visibility data at full spectral resolution. 

In the high S/N regime, an important aspect of achieving high dynamic range is a reliable method for accounting for the spectral behaviour of the continuum. To this end, we adapted the method used by A15 to improve the accuracy of the optical depth measurement. First, the \texttt{CLEAN}
deconvolution algorithm was used to construct a continuum model of the
field in each 4 and 5\,MHz beam-forming interval. The target quasar was then masked and the model subtracted from the corresponding visibility data, effectively removing other bright sources
in the field. We then used an adapted version of the \texttt{UVLIN}
task to fit a second-order polynomial to the continuum, which was then divided out and a value of unity subtracted to form normalized continuum-subtracted
visibilities. Importantly, this adapted method
accounted for any gradient and/or curvature in the
continuum\footnote{Frequency-dependent features in the continuum can
  either be generated by intrinsic structure in the source spectrum, 
  residual source-confusion or the instrumental band-pass. For
  example, we see a low-level 25\,MHz-period ripple towards the
  low-frequency end of the band, which is caused by a standing wave
  between the dish and focal plane.} which would contribute systematic error to our estimate of the optical depth in high S/N data. Imaging of the visibilities was carried out using the standard procedure with natural weighting to optimize the S/N for
detecting absorption lines. We identified the centroid position of the quasar from the peak flux density in the averaged continuum image, and then extracted the spectra at this position from each normalized continuum-subtracted cube using \texttt{MBSPECT}. 

\subsection{Spectral analysis}\label{section:analysis}

We used three frequency bands in our observations, enabling a search
for \mbox{H\,{\sc i}} absorption spanning redshifts between $z =
0.0$ and
$1.0$ (or $z = 0.1$ and $1.3$ for the 18\,cm OH lines). For each of
our target quasars we generated a single composite
spectrum by shifting the individual spectra to the common Solar barycentre rest frame and taking the variance-weighted arithmetic mean. The composite ASKAP BETA spectra are shown in
\autoref{figure:PKS1830-211_beta_spectrum} and
\autoref{figure:MGJ0414+0534_beta_spectrum}, where we have indicated
those 21-cm \mbox{H\,{\sc i}} and 18-cm OH lines previously reported
in the literature. Approximately 14\,per\,cent of the redshift
coverage was rendered unavailable due to satellite and aviation
generated RFI, predominantly at $z \lesssim 0.4$. We used the spectral line detection method
discussed by \cite{Allison:2012b, Allison:2014} to identify absorption
lines in each spectrum, which performs multimodal nested sampling
(\citealt{Feroz:2008}; \citealt{Feroz:2009b}) to calculate the probability of spectral lines in the data.

\begin{figure}
\centering
\includegraphics[width=0.445\textwidth]{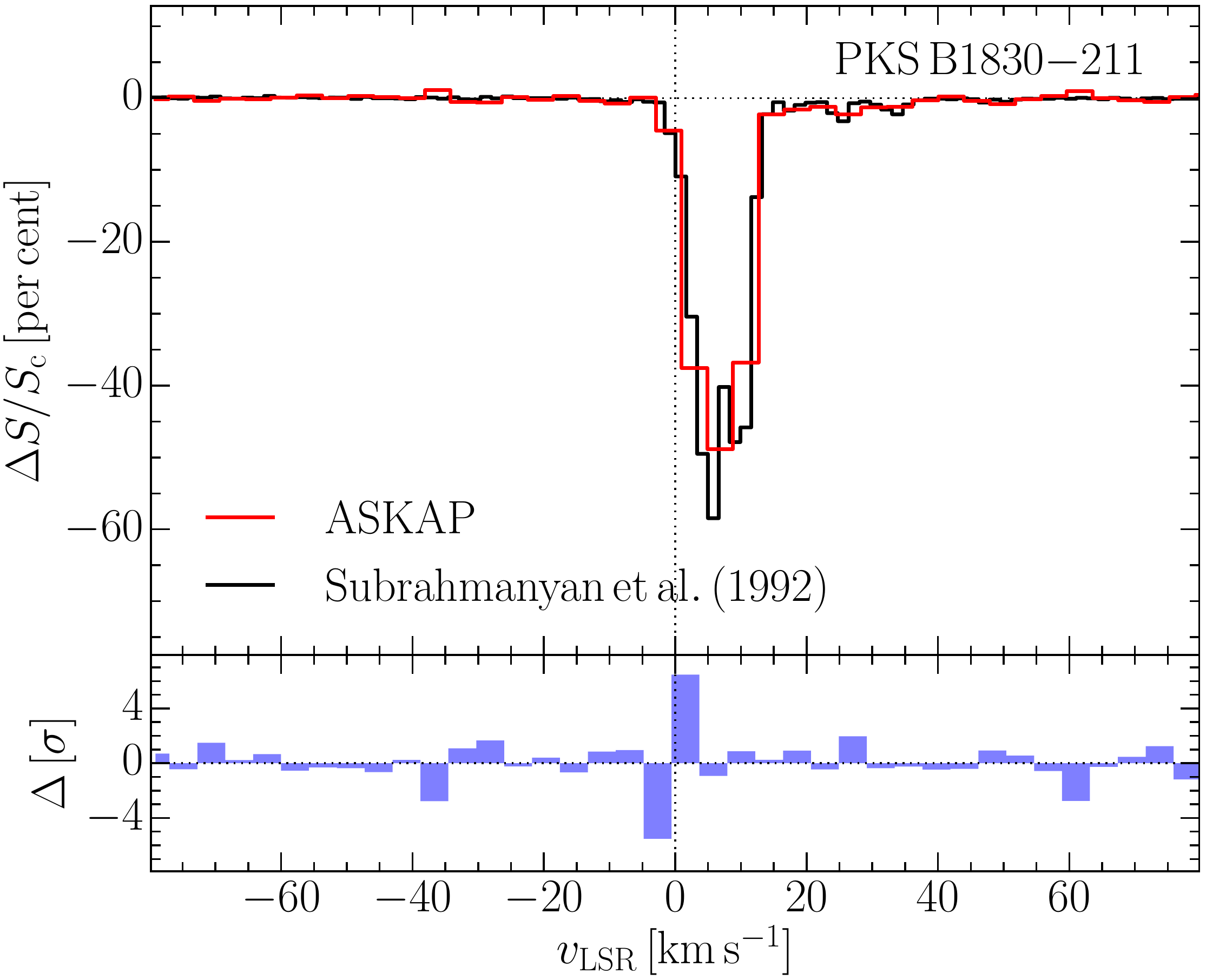}
\caption{Spectra from ASKAP BETA (this work) and ATCA
  (\citealt{Subrahmanyan:1992}) showing Galactic \mbox{H\,{\sc i}}
  absorption towards PKS\,B1830$-$211. The difference
  spectrum (blue filled) is shown in the bottom
  panel. Although $\sim$6\,$\sigma$ peaks are apparent in the difference spectrum, the line is not sufficiently sampled by the 18.5\,kHz ASKAP BETA channels and we attribute these features to artefacts of the binning rather than temporal changes in the line-of-sight gas.}\label{figure:PKS1830-211_hi_mw_comparison}
\end{figure}

In several figures throughout this paper
(e.g. \autoref{figure:PKS1830-211_hi_mw_comparison}), we compare our
spectra with those in the literature, showing in each case the
difference spectrum normalized by the rms noise. In order to account
for the different resolutions and frequency sampling in each
observation, we have smoothed and re-sampled (using cubic spline
interpolation) each spectrum to a common resolution which is equal to the
quadrature sum of the two individual resolutions. This process ensures
that the frequency bins in the resulting difference spectrum are
independent and can therefore be used to identify any significant
deviations from zero. In the case of ASKAP BETA the 
response of the 18.5\,kHz channels (\citealt{Tuthill:2012}) produces a correlation of only $\rho_{\rm chan} \approx
4\times10^{-4}$ between adjacent channels and so we can assume a spectral resolution
equal to the channel spacing. For the other spectra taken
from the literature, we have used the resolution information available from those publications.

\section{PKS\,B1830--211}\label{section:pks1830-211}

PKS\,B1830$-$211 is a well-studied radio-loud blazar at a redshift of $z
= 2.51$ (\citealt{Lidman:1999}) which is strongly gravitationally
lensed by a foreground galaxy (e.g. \citealt{Subrahmanyan:1990, Jauncey:1991, Nair:1993}). At cm-wavelengths, the image comprises two compact flat-spectrum components, separated by approximately 1\,arcsec and orientated to the north-east (NE) and south-west (SW) of the lens centre (e.g. \citealt{Rao:1988, Subrahmanyan:1990}), which are connected by an Einstein ring of steep-spectrum emission from the lensed jet (\citealt{Jauncey:1991}). The spiral galaxy lens at $z = 0.886$ (\citealt{Wiklind:1996}) is rich in molecules, with over 40 different species detected so far, the majority of which are seen in absorption against the compact SW image of the core (see e.g. \citealt{Muller:2014} and references therein). \mbox{H\,{\sc i}} absorption was also previously detected in both this (\citealt{Chengalur:1999, Koopmans:2005}) and a possible second intervening galaxy at $z = 0.192$ (\citealt{Lovell:1996}).

The flux density of this source at cm-wavelengths ($\sim$10\,Jy) makes it a perfect target for a line-of-sight survey for 21-cm \mbox{H\,{\sc i}} and 18-cm OH absorption. Using commissioning data from ASKAP BETA, we have detected a total of three intervening systems between $z_{\rm HI} = 0$ and 1, consisting of \mbox{H\,{\sc i}}
absorption in the Milky Way, \mbox{H\,{\sc i}}, and OH absorption in the $z = 0.886$ galaxy, and probable \mbox{H\,{\sc i}} absorption in a second intervening galaxy at $z = 0.192$. All three systems were previously reported in the literature and we detected no further intervening galaxies throughout the spectrum shown in \autoref{figure:PKS1830-211_beta_spectrum}. In the following, we discuss each of these systems in more detail.

\subsection{Absorption in the Galactic ISM}

Strong \mbox{H\,{\sc i}} absorption ($\tau_{\rm peak} \approx
0.9$) is seen at 1420\,MHz along the line of sight to PKS\,B1830$-$21, 
which passes within 2\,kpc of the Galactic Centre and leaves the Solar circle
on the far side at about 2\,kpc below the Galactic plane ($b = -5\fdg7, l = 12\fdg2$).
Originally reported by \cite{Subrahmanyan:1992} using the Australia Telescope
Compact Array (ATCA), several distinct velocity components can be seen in the line extending from approximately $-$15 to 100\,km\,s$^{-1}$ in the local standard of rest (LSR) velocity frame, tracing cold atomic gas from within 17\,kpc of the Sun
(positive velocities) to beyond the Solar circle
(negative velocities). By identifying that the weak $-$9\,km\,s$^{-1}$
component was common to  both the emission and absorption lines, \cite{Subrahmanyan:1992} argued that PKS\,B1830$-$211 was likely located beyond the \mbox{H\,{\sc
i}} disc of the Milky Way and therefore extragalactic in origin. Notwithstanding that
our ASKAP BETA data have lower spectral resolution (by a factor of
approximately 2.5), we find that the bulk of this velocity structure is also seen in our data (\autoref{figure:PKS1830-211_hi_mw_comparison}).

Over the 23\,yr interval spanned by the ASKAP BETA and ATCA data it is possible that temporal variations in the line profile could arise from the motion of individual optical depth fluctuations transiting the quasar sight line. For example, in the case of absorbing gas on the far side of the Galactic Solar circle,
we expect transverse velocities of up to $\sim$400\,km\,s$^{-1}$
with respect to the quasar sight line (e.g. \citealt{Reid:2014}), corresponding to a physical scale of $\sim$2000\,au. At a distance of $\sim$17\,kpc from the Sun this physical scale would
equate to angular scales of $\sim$0.1\,arcsec. Although the entire extent of the source at cm-wavelengths is about 1\,arcsec, at least 10\,per\,cent of the flux density is contained within $\sim$0.1\,arcsec each of the compact NE and SW images of the quasar core (\citealt{Jauncey:1991, Lovell:1996}). Since the quasar sight-line passes near to the Galactic Centre, images of the core at cm-wavelengths are significantly scatter broadened by the ISM (\citealt{Jones:1996}) and their intrinsic sizes are likely to be smaller than this (e.g. \citealt{Garrett:1997}). The NE and the SW components are therefore sufficiently compact at cm-wavelengths to resolve 21-cm optical depth fluctuations on 2000\,au scales at 17\,kpc. Certainly transverse structures with significant 21-cm optical depths ($\Delta{\tau} \sim 0.1 - 1$) have previously been reported on 10 -- 1000\,au scales (e.g. \citealt{Johnston:2003, Brogan:2005, Goss:2008, Lazio:2009}), and could arise naturally from a power spectrum of turbulent fluctuations in the ISM (e.g. \citealt{Deshpande:2000b, Roy:2010, Roy:2012, Dutta:2014}). 

In \autoref{figure:PKS1830-211_hi_mw_comparison}, we show a comparison of our spectrum with 
that of \cite{Subrahmanyan:1992}, including the
difference shown in the bottom panel (in units of the rms noise). The noise per 
18.5\,kHz channel in the ASKAP BETA spectrum is approximately a factor of 5
times that of the ATCA spectrum (at the same resolution) and is the
dominant contributor to the rms noise in the difference spectrum. Since
\cite{Subrahmanyan:1992} measured a peak \mbox{H\,{\sc i}}
brightness temperature of approximately 50\,K, we expect the noise to be
higher on the line due to the contribution of \mbox{H\,{\sc i}}
emission to the system temperature (c.f. the 120\,K system temperature expected off the line for ASKAP BETA; \citealt{McConnell:2016}). However, we account for this by normalizing the difference spectrum based on the measured
rms noise per channel. We see two features either side of $v_{\rm
LSR} = 0$\,km\,s$^{-1}$ which appear significant at a level $\sim6\,\sigma$ and
correspond to a change in 21-cm optical depth of $\Delta{\tau} \sim 0.05/c_{\rm f}$, where $c_{\rm f}$ is the covering factor of the whole source. In the case of the NE and the SW images of the  core, these spectral features would then correspond to optical depth fluctuations equal to $\Delta{\tau} \sim 0.5$ (i.e. $c_{\rm f} \sim 0.1$). Unfortunately, the absorption line is not well sampled by either the ASKAP BETA or ATCA data and so the difference spectrum is very sensitive to the re-sampling processes used for comparison. We find that the  apparent 6\,$\sigma$ differences seen between the spectra are easily affected by changes in the binning. However, we suggest that future monitoring of this sight line at higher spectral resolution with ASKAP (see e.g. \citealt{Dickey:2013}) could potentially reveal optical depth variations due to transverse motion of the small-scale Galactic CNM.

\begin{figure*}
\centering
\includegraphics[width=1.0\textwidth]{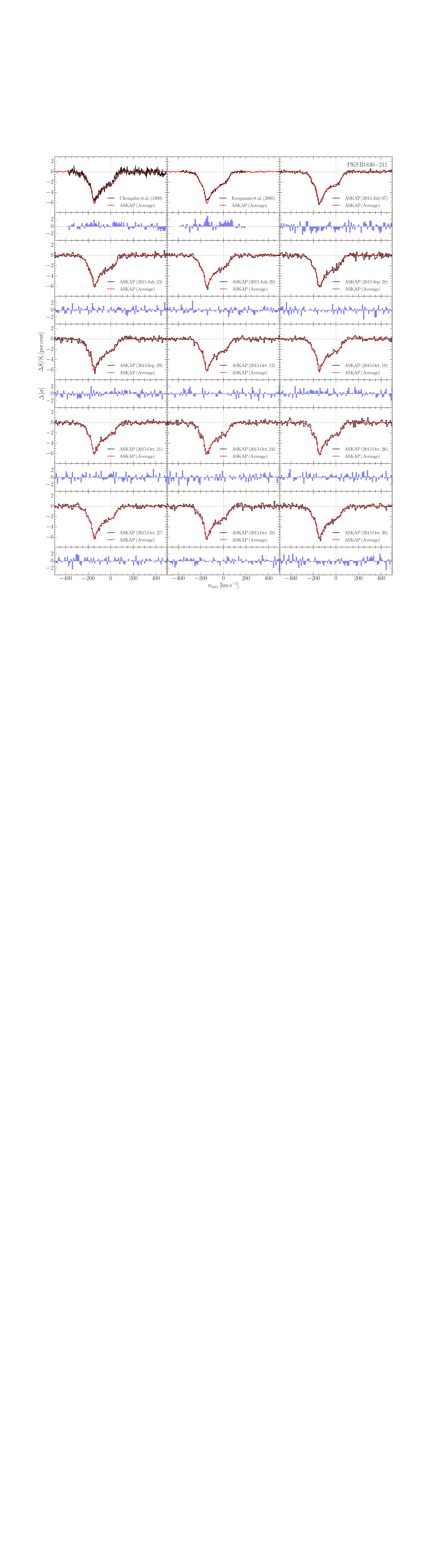}
\caption{Spectra showing \mbox{H\,{\sc i}} absorption arising in the face-on spiral 
	galaxy at $z = 0.88582$ towards
  PKS\,B1830$-$211. The red line in each panel is a variance-weighted
  arithmetic mean of the data from all observations with the ASKAP BETA telescope over the period 2014 -- 2015. A total of 15 epochs are shown,
  including two previous observations using the WSRT in 1997
  (\citealt{Chengalur:1999}) and 1998 (\citealt{Koopmans:2005}). The
  difference spectra (blue filled) are shown in the panels
  below each epoch.}\label{figure:PKS1830-211_hi_z089_comparison}
\end{figure*}

\begin{figure}
\centering
\includegraphics[width=0.45\textwidth]{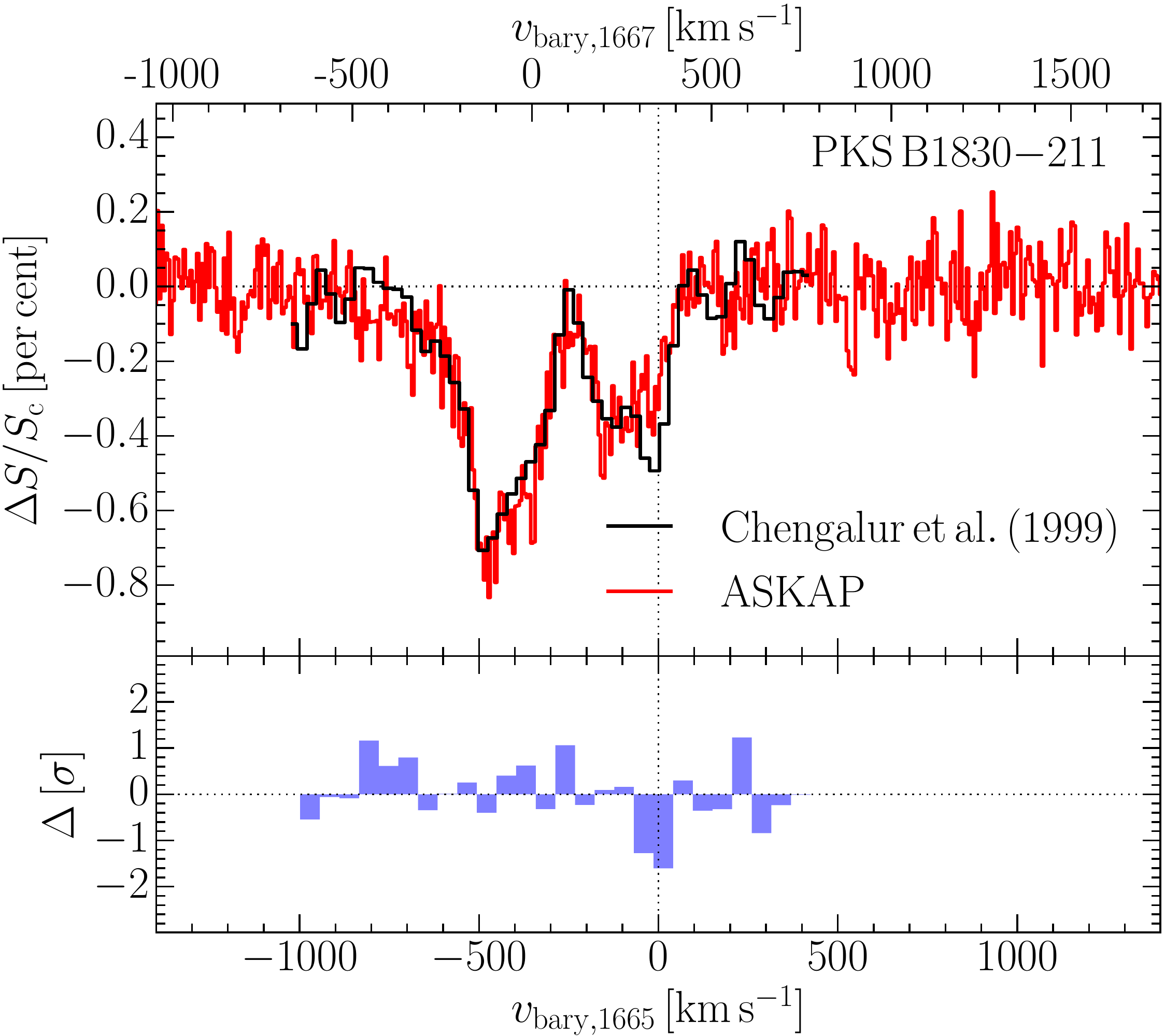}
\caption{Spectra from ASKAP BETA (this work) and WSRT
  (\citealt{Chengalur:1999}) showing the 1665 and 1667\,MHz
OH absorption lines at $z = 0.88582$ towards PKS\,B1830$-$211. The lower
and upper velocity scales are given for the 1665 and 1667\,MHz lines,
respectively. The difference spectrum (blue filled) is shown in the
bottom  panel.}\label{figure:PKS1830-211_oh_z089_comparison}
\end{figure}

\subsection{Intervening galaxies}

\subsubsection{The spiral galaxy lens at $z = 0.886$}

So far more than 40 high-redshift absorption lines have been detected in the radio and sub-mm spectrum towards PKS\,B1830$-$211 due to an intervening galaxy at $z = 0.886$ (see \citealt{Muller:2014} and references therein). The spatial extent and velocity distribution of the gas inferred from absorption against the lensed structure of the source are consistent with a massive early-type spiral galaxy which is near to face on with respect to the line of sight (\citealt{Wiklind:1998, Chengalur:1999, Koopmans:2005}). This interpretation is supported by optical
imaging using the \emph{HST} which confirmed a spiral galaxy with clearly delineated spiral arms (\citealt{Courbin:2002, Winn:2002}). One of the spiral arms appears to intercept the SW core image where several saturated molecular absorption lines are detected (e.g. \citealt{Frye:1997, Muller:2011}).   

In \autoref{figure:PKS1830-211_hi_z089_comparison}, we show the \mbox{H\,{\sc i}} absorption line associated with this galaxy. Two distinct velocity components are evident and can be equated with gas at the positions of the NE ($-150$\,km\,s$^{-1}$) and SW (0\,km\,s$^{-1}$) core images (e.g. \citealt{Chengalur:1999}). The full width of the line spans almost 400\,km\,s$^{-1}$ and is consistent with a low-inclination \mbox{H\,{\sc i}} disc obscuring the entire image of the radio source, including the steep spectrum Einstein ring seen at 21-cm wavelengths. The inferred column density of $N_{\rm HI} \gtrsim 2 \times 10^{21}$\,cm$^{-2}$ (assuming $T_{\rm spin} \gtrsim 100$\,K) is consistent with the disc of a large dusty spiral galaxy. By simultaneously fitting the 21-cm spectrum and a spatially resolved VLBI image of the continuum, \cite{Koopmans:2005} showed that the absorption line could be modelled using a constant-velocity axisymmetric rotational disc with a radially dependent 21-cm optical depth. They estimated an inclination angle ($i \approx 17\degr - 32\degr$) which is consistent with the orientation inferred from the earlier radio and optical data, and a velocity dispersion ($\Delta{v} \approx 39 - 48$\,km\,s$^{-1}$) which is similar to that measured for the molecular lines (e.g. \citealt{Wiklind:1996}). The approximately 2:1 ratio of the NE and the SW components in the line is consistent with an optical depth which increases with galacto-centric radius and the closer proximity of the SW core to the \mbox{H\,{\sc i}}-depleted centre (constrained by lens models; e.g. {\citealt{Kochanek:1992, Nair:1993}). 

Using the ASKAP BETA telescope, we monitored this line for 13 epochs during 2014 and 2015 in order to test for variability over the period spanned by our observing campaign and previous observations undertaken with the Westerbork Synthesis Radio Telescope (WSRT) almost 20\,yr earlier (\citealt{Chengalur:1999, Koopmans:2005}). In \autoref{figure:PKS1830-211_hi_z089_comparison}, we compare each epoch to the average ASKAP BETA spectrum, and find that the \mbox{H\,{\sc i}} absorption appears to vary on time-scales of approximately 1\,yr. We discuss these results further in \autoref{section:monitoring_results}, but note here that the stability of the ASKAP BETA spectra obtained for all observing epochs during 2014 and 2015 is very encouraging for future \mbox{H\,{\sc i}} absorption surveys with this telescope. 
 
We also detect both of the 1665 and 1667\,MHz main doublet lines of OH in absorption (see \autoref{figure:PKS1830-211_oh_z089_comparison}), originally reported by \cite{Chengalur:1999}. The line strength is approximately in the 9:5 ratio expected from optically thin thermalized OH. If we assume a constant excitation temperature then we would also expect to see the 1612 and 1721\,MHz satellite lines at an optical depth of approximately 0.08\,per\,cent. At the redshifted frequencies of these lines (855 and 912\,MHz), the rms optical depth sensitivity of our data per 18.5\,kHz channel is approximately 0.07\,per\,cent. Assuming a line width of $\sim$50\,km\,s$^{-1}$, our data would be sensitive to the satellite lines at the level $\sim$3$\,\sigma$ and so non-detection is probably still consistent with thermalized OH, but future deeper observations should provide either detections or stronger constraints on departure from this assumption. The widths of the main lines suggest that the OH is widespread throughout the galactic disc, covering continuum emission from both the NE and the SW core images and the Einstein ring. We find no significant evidence for optical depth variation between the 1665 and 1667 lines suggested by \cite{Chengalur:1999}; both lines appear to have higher absorption by a factor $\sim$2 at -150\,km\,s$^{-1}$ compared with $0$\,km\,s$^{-1}$. This is similar to the ratio for the \mbox{H\,{\sc i}} components and different to the other molecular absorption lines in this galaxy (e.g. \citealt{Muller:2011}), where absorption against the NE image of the core is typically weaker than its SW counterpart. A likely scenario is that the OH absorption is tracing a more diffuse molecular component than spectral-line tracers of denser gas seen at mm-wavelengths. No significant variability is detected in the OH line, over the 20\,yr time interval spanned by the ASKAP BETA and WSRT data, although the S/N is significantly lower than that of the \mbox{H\,{\sc i}} line. 

\begin{figure}
\centering
\includegraphics[width=0.45\textwidth]{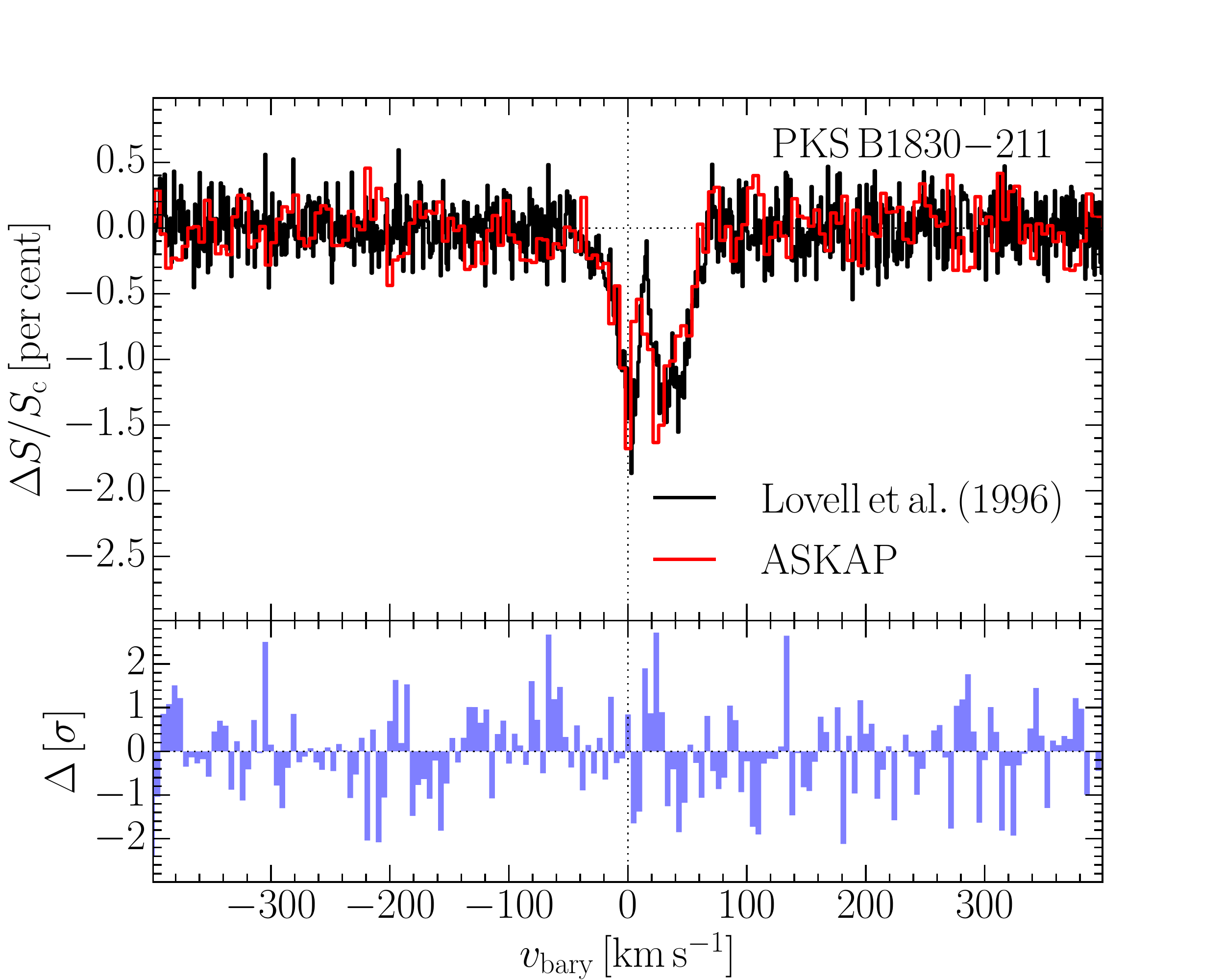} \\
\caption{Spectra from ASKAP BETA (this work) and ATCA
  (\citealt{Lovell:1996}) showing \mbox{H\,{\sc i}} absorption
  associated with the possible intervening galaxy at $z = 0.1926$ towards
  PKS\,B1830$-$211. The difference spectrum (blue filled)
  is shown in the bottom
  panel.}\label{figure:PKS1830-211_hi_z019_comparison}
\end{figure}

\subsubsection{A possible second galaxy at $z = 0.193$}

A further absorption line at 1190\,MHz was discovered by \cite{Lovell:1996}
using the Parkes 64-m radio telescope, which they then
confirmed following observations using the Australia Telescope Compact
Array (ATCA) and four antennas of the Australian Long Baseline Array
(LBA). Based on the relatively strong absorption against the NE image, in contrast to molecular absorption lines detected at $z = 0.886$, and the lack of common transitions known to exist at 2246\,MHz (but see also \citealt{Wiklind:1998}), they argued that this line was associated with a second intervening \mbox{H\,{\sc i}} absorber at a redshift
of $z = 0.193$. The implied column density $N_{\rm HI} \sim 10^{21}$\,cm$^{-2}$ (assuming $T_{\rm spin} \approx 100$\,K; \citealt{Frye:1997}) is typical of a large spiral galaxy. A possible optical counterpart was found in the \emph{HST} images
of PKS\,B1830$-$211 by \cite{Courbin:2002} and \cite{Winn:2002}, providing further evidence of a second intervening spiral galaxy. However, there is yet to be spectroscopic confirmation at either optical or 
sub-mm wavelengths (\citealt{Muller:2011}). With ASKAP BETA we confirm the existence of the 1190\,MHz line and, at a 5\,$\sigma$ optical depth sensitivity of 0.16\,per\,cent  (assuming a line width of $\sim$50\,km\,s$^{-1}$), the absence of any corresponding absorption
line at 1016\,MHz (\autoref{figure:PKS1830-211_beta_spectrum}), supporting the argument that
the 1190\,MHz line is more likely to be \mbox{H\,{\sc i}} than OH. At 1396 and 1398\,MHz (where one might expect to detect the main OH doublet lines), the 5\,$\sigma$ optical depth sensitivity of our data is only 0.46\,per\,cent, giving a decidedly non-informative upper limit of $\sim$3 for the \mbox{H\,{\sc i}} to OH ratio. We find good agreement between the line profile in the ASKAP BETA spectrum and that of \cite{Lovell:1996}, recovering the two distinct components associated with gas (possibly in $\sim$\,kpc spiral arms) which cover the NE core and Einstein ring components of the radio source (\autoref{figure:PKS1830-211_hi_z019_comparison}). At the S/N limit of the data, we find no evidence for any changes in the spectrum over the 20\,yr time interval between these observations.

\subsection{Spectral variability in the gravitational
  lens}\label{section:monitoring_results}

Upon visual inspection of the difference spectra, shown in \autoref{figure:PKS1830-211_hi_z089_comparison}, it is apparent that the \mbox{H\,{\sc i}} line in the $z = 0.886$ absorber towards PKS\,B1830$-$211 varies significantly above the noise. This is particularly clear when we compare the average ASKAP BETA spectrum with two published spectra from previous
observations undertaken with the WSRT in 1997 (\citealt{Chengalur:1999}) and
1998 (\citealt{Koopmans:2005})\footnote{Note that we have corrected
  the baseline for the \citet{Koopmans:2005} spectrum by fitting and
  subtracting a linear component. We found that the  baseline in the published 
  spectrum appeared to deviate from zero by up to 0.13\,per\,cent of the continuum.}. \cite{Koopmans:2005} also noted seeing differences between the WSRT spectra, but as yet these have not been published. Absorption line variability can be used to infer the scale of optical depth variations in foreground absorbing gas not normally possible with conventional imaging techniques. Mechanisms driving variability on day to year time-scales arise either from the apparent superluminal motion of beamed jet-knots in the background source (e.g. \citealt{Briggs:1983}), ISS of either sufficiently compact components of the continuum source or the absorbing region (e.g. \citealt{Macquart:2005}), or microlensing by small-scale changes in the foreground mass distribution (e.g. \citealt{Lewis:2003}). It is plausible that changes in the physical properties of the absorbing gas can also drive variability. For example, \cite{Wolfe:1982} claimed that if the excitation (spin) temperature is coupled to a nearby source of 21-cm continuum emission, which varies strongly, then detectable variation in the 21-cm optical depth would be possible. 

Only a few examples of \mbox{H\,{\sc i}} absorption variability have been reported in the literature so far. Notably, \cite{Wolfe:1982} found significant changes in four narrow absorption components in the 21-cm spectrum of BL Lacertae object AO\,0235$+$164, with durations of several months. Although the original data gave contradictory evidence in support of either intrinsic or extrinsic source models for the variability (see also \citealt{Briggs:1983}), ultimately subsequent optical spectroscopic confirmation of the source redshift (\citealt{Cohen:1987}) conclusively favoured the latter scenario. A second variable 21-cm absorber was found by \cite{Kanekar:2001a} in the DLA system towards the quasar PKS\,B1127$-$145, with a characteristic time-scale of a few days. Since there was no apparent correlation in the source and line flux densities, the authors favoured models with transverse motion of a knot in the background quasar, which possibly scintillates from either the Galactic and/or absorber ISM. The spectral variability which we see for the \mbox{H\,{\sc i}} absorber in the PKS\,B1830$-$211 lens is therefore a rare example of 21-cm variability and its unique nature as a strongly lensed source may reveal new behaviour.

\subsubsection{Modeling the spectral variability}

By parametrizing spectral variability in terms of the equivalent width (EW), we can separate intrinsic changes in the optical depth and geometry of the absorber-source system from simple systematic errors or variations in the absolute flux density. In \autoref{figure:time_series} we show, as a function of observing
epoch, the total cm-wavelength continuum flux density, the 
EW of the whole \mbox{H\,{\sc i}} line, the NE and the SW components, and their ratio. We avoid possible model-dependent artefacts which could result from fitting analytical profiles by measuring EW directly from the data shown in \autoref{figure:PKS1830-211_hi_z089_comparison}, over velocity channels spanning the absorption line. To detect any relative changes in absorption across the lensed radio image, we also consider absorption separately against the bright NE and SW components of the core by integrating sections of the line centred at -150 and 0\,km\,s$^{-1}$ (with respect to $z = 0.88582$) over a velocity width of $\Delta{v} = \pm$40\,km\,s$^{-1}$, equal to the line broadening given by the model of \cite{Koopmans:2005}. 

\begin{figure*}
\centering
\includegraphics[width=1.0\textwidth]{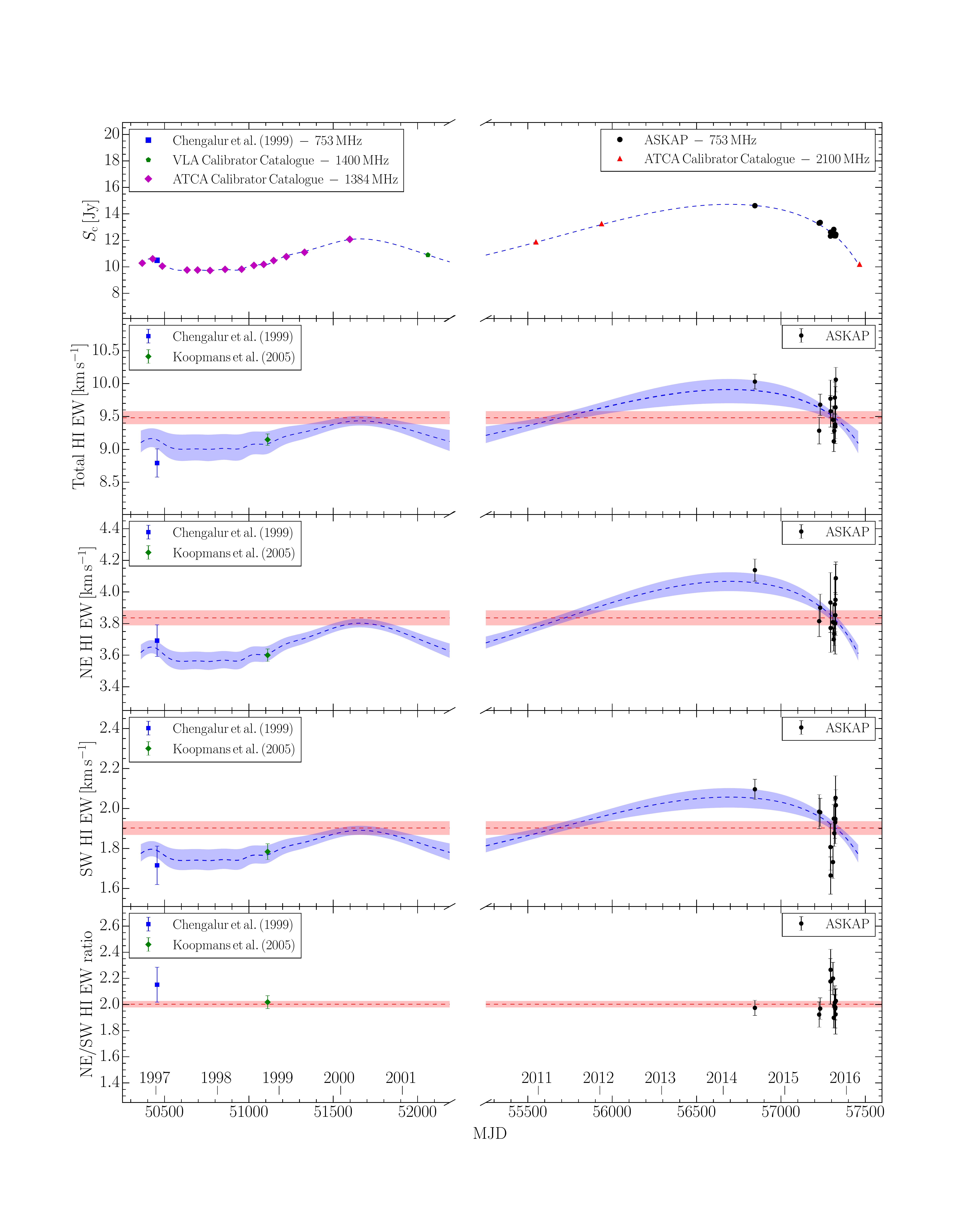}
\caption{Time-series data for the total source flux density (top
  panel), EWs of the \mbox{H\,{\sc i}} line, NE and SW
  components (middle panels), and their ratio (bottom panel), for the intervening spiral galaxy at $z = 0.88582$ towards PKS\,B1830$-$211. The results of model fitting are denoted by dashed lines (the median) and filled regions (68.3\,per\,cent credible interval), where red models represent constant behaviour with respect to time and blue models are based on a cubic spline interpolation in the continuum data. The blue models are fit to the EW data using an offset and scaling parameter.}\label{figure:time_series}
\end{figure*}

\begin{figure*}
\centering
\includegraphics[width=1.0\textwidth]{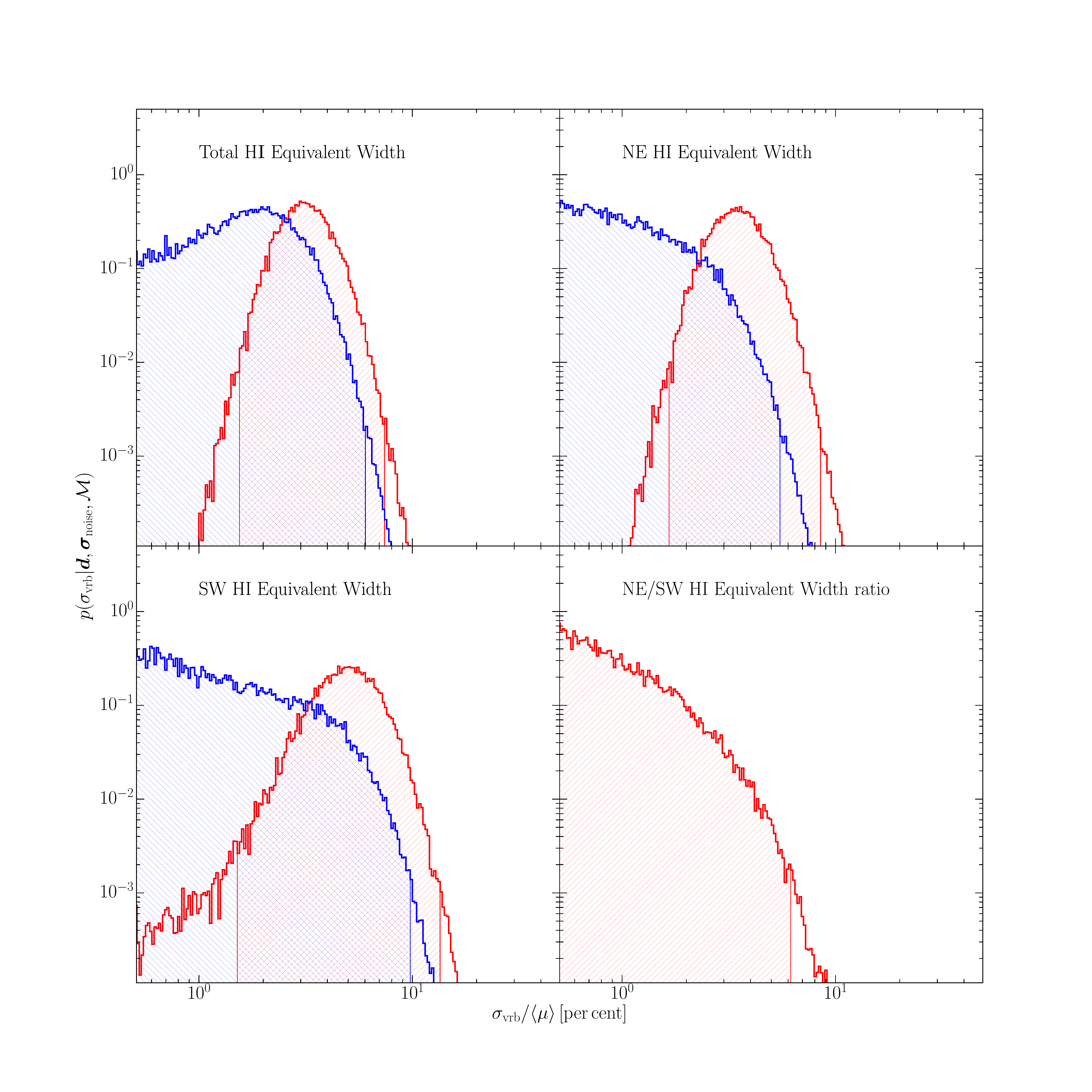}
\caption{Marginal posterior probability distributions for fractional residual variability ($\sigma_{\rm vrb}/\langle\mu\rangle$) in the EWs of the \mbox{H\,{\sc i}} line, the NE 
and the SW components, and their ratio. The bounded hatched regions
  represent the 99.7\,per\,cent (3\,$\sigma$) credible interval about the
  distribution median. The red and blue distributions correspond to the two models ($\mathcal{M}$) shown in \autoref{figure:time_series}.}\label{figure:variability_probability}
\end{figure*}

We employ a simple quantitative test by modelling the residual variability as a purely stochastic process which is independent between time samples and has a normal distribution of standard deviation $\sigma_{\rm vrb}$. Although the assumption of independence is weak, we can nevertheless test for any residual deviation from a given temporal model ($\mathcal{M}$), above that expected from just the noise\footnote{This approach has been used for other applications, see for example \citet{Purcell:2015}}. The likelihood of $\sigma_{\rm vrb}$ assuming a normal distribution for the data is then given by
\begin{align}
& \mathcal{L}(\sigma_{\rm vrb}) \equiv p(\boldsymbol{d}| \sigma_{\rm vrb}, \boldsymbol{\sigma}_{\rm noise}, \mathcal{M}), \nonumber \\ 
& = \mathlarger{\int} {\pi(\boldsymbol{\theta})\over\sqrt{(2\mathrm{\pi})^{N}\prod_{i}^{N}{\sigma_{{\rm total},i}^2}}}\exp{\left[-{1\over2}{\sum_{i}^{N}{(d_{i} -\mu_{i}(\boldsymbol{\theta}))^{2}\over\sigma_{{\rm total},i}^2}}\right]}\mathrm{d}\boldsymbol{\theta},
\end{align}
where
\begin{equation}
\sigma_{{\rm total},i}^{2} = {\sigma}_{{\rm noise},i}^{2} + \sigma_{\rm vrb}^{2},
\end{equation} 
$\boldsymbol{d}$ is the set of measured data $[d_{i}, ..., d_{N}]$, $\boldsymbol{\mu}$ are the
expected values $[\mu_{i}, ..., \mu_{N}]$ assuming a model $\mathcal{M}$ defined by parameters $\boldsymbol{\theta}$ with prior probability $\pi(\boldsymbol{\theta})$, and $\boldsymbol{\sigma}_{\rm noise}$ are the expected standard deviations $[\sigma_{{\rm noise},i}, ..., {\sigma}_{{\rm noise},N}]$ estimated from the measured rms noise in each observation. We can then compute the posterior probability density $p(\sigma_{\rm vrb}|\boldsymbol{d},\boldsymbol{\sigma}_{\rm noise},\mathcal{M})$ by taking the product of $\mathcal{L}(\sigma_{\rm vrb})$ and a suitable prior density function $\pi(\sigma_{\rm vrb})$. We assume a\,priori ignorance of the value of $\sigma_{\rm vrb}$ and adopt the following scale-invariant Jeffreys prior
\begin{equation}
  \pi(\sigma_{\rm vrb})= 
\begin{cases}
  [3\ln(10)\,\sigma_{\rm vrb}]^{-1}, & \text{if}\ 10^{-3} \leq \sigma_{\rm vrb}/\langle{\mu}\rangle < 1, \\
  0, & \text{otherwise}.
\end{cases}
\end{equation}
where we scale to $\langle{\mu}\rangle$, the mean of our data model. For the general class of models $\mathcal{M}$ we estimate $p(\sigma_{\rm vrb}|\boldsymbol{d},\boldsymbol{\sigma}_{\rm noise},\mathcal{M})$ and its integral (the Bayes factor) using \texttt{PyMultiNest} (\citealt{Buchner:2014}), which implements the \texttt{MultiNest} algorithm (\citealt{Feroz:2008,Feroz:2009b,Feroz:2013}). 

In the first instance, we consider a model which is time independent so that all $\mu_{i} = \langle{\mu}\rangle$, the maximum likelihood of $\langle{\mu}\rangle$ simply being equal to the noise-weighted mean of the data. We show in \autoref{figure:time_series}} the median and 68.3\,per\,cent credible interval for $\langle{\mu}\rangle$ and in \autoref{figure:variability_probability} the marginalized posterior probability distribution for $\sigma_{\rm vrb}/\langle{\mu}\rangle$. We calculate 99.7\,per\,cent (3\,$\sigma$) credible intervals for $\sigma_{\rm vrb}/\langle{\mu}\rangle$ of $1.5$ -- $7.5$\,per\,cent in the total EW, $1.7$ -- $8.5$\,per\,cent in the NE component, and $1.5$ -- $13.4$\,per\,cent in the SW component. By integrating the marginal posterior probability distribution over all $\sigma_{\rm vrb}/\langle{\mu}\rangle$, and dividing by the posterior for the noise-only model, we calculate variability-to-noise-only Bayes factors of $\sim$$3\times10^{7}$ in the total EW, $\sim$$1\times10^{7}$ in the NE component, and $\sim$$1\times10^{3}$ in the SW component, equivalent to two-model probabilities of 99.999997, 99.999991, and 99.91\,per\,cent. Although there is clearly overwhelming evidence for variability in the EW, above that expected from just the measured noise, we find no evidence of variability in the ratio of the NE and SW components; $\sigma_{\rm vrb}/\langle{\mu}\rangle$ is consistent with zero and has a 99.7\,per\,cent (3\,$\sigma$) upper limit of 6\,per\,cent, a Bayes factor of 0.38 and a two-model probability of 28\,per\,cent. So while the \mbox{H\,{\sc i}} EW varies at the level of a few percent on yearly time-scales, it does so in a manner that is apparently coherent between the NE and the SW components. This implies that the variability we see in the \mbox{H\,{\sc i}} line is coupled to intrinsic brightening and fading of the core rather than any ISS or microlensing, which should produce independent variability in these two components.  

We verify this interpretation by testing for a positive correlation between temporal variations in the total continuum flux density and the EWs of the line components. We construct a second model based on a cubic spline interpolation of the temporal behaviour in the total continuum flux density (assuming a spectral index of approximately zero over these frequencies), and then fit this to the \mbox{H\,{\sc i}} EW data with a linear scaling and offset. The results are shown in \autoref{figure:time_series} and we obtain linear scalings between the continuum and EWs of the total \mbox{H\,{\sc i}}, NE, and SW components equal to  $0.24_{-0.11}^{+0.08}$, $0.33_{-0.07}^{+0.06}$, and $0.42_{-0.13}^{+0.11}$, respectively. The marginal posterior probability distributions for any residual variability are shown in \autoref{figure:variability_probability} and we find no evidence for residual variability in either the NE or SW components of the line, with probabilities of 36 and 38\,per\,cent, respectively. There is weak evidence for variability in the total EW with a probability of 77\,per\,cent, which may indicate that a more complex model beyond a simple linear relationship is required, or that further variability is being driven by another process such as ISS.

Correlated variability naturally arises between the background flux density and EW of the \mbox{H\,{\sc i}} line as a direct consequence of changes in the source brightness distribution. Consider an element of optical depth $\tau_{i}$ and velocity width $\Delta{v}_{i}$, obscuring a component $S_{i}$ of the background continuum $S_{\rm c}$. The EW of this element is then given by
\begin{equation}
\mathrm{EW}_{i} \approx \frac{S_{i}}{S_{\rm c}}\left(1-{\rm
e}^{-\tau_{i}}\right)\Delta{v}_{i}.
\end{equation}
If we assume that the optical depth does not change significantly on the time-scales sampled by our data, then we derive the following relationship between fractional changes in the EW and the total continuum
\begin{equation}
\frac{\delta{\mathrm{EW}}_{i}}{\mathrm{EW}_{i}} \approx \left({\frac{f_{i}^{\prime}}{f_{i}} - 1}\right)\frac{\delta{S}_{\rm c}}{S_{\rm c}},
\end{equation}
where $f_{i} = S_{i}/S_{\rm c} \leq 1$ and $f_{i}^{\prime} = \delta{S}_{i}/\delta{S}_{\rm c} \leq 1$. Fractional changes in $\mathrm{EW}_{i}$ and $S_{\rm c}$ are positively correlated if $f_{i}^{\prime} > f_{i}$, i.e. the source preferentially brightens or fades in this component. In the case of PKS\,B1830$-$211, we find that  fractional changes in the $\mathrm{EW}$s which are nominally associated with the NE and the SW core images are $\sim$30\,per\,cent that of the total continuum, from which we can infer that no more than $\sim$75\,per\,cent of the total continuum is contained in these components (consistent with imaging). The inequality comes about because we cannot, in practice, resolve individual components in the \mbox{H\,{\sc i}} absorption line (which have a typical width $\sim$40\,km\,s$^{-1}$; \citealt{Koopmans:2005}), and so there will be some contribution from non-varying components of the source. Under this model, we also expect the steep-spectrum Einstein ring not to vary on these time-scales (i.e.  $f^{\prime}_{i} = 0$) and so fractional changes in $\mathrm{EW}$ for this component should be 100\,per\,cent anti-correlated with the total continuum flux density. The overall affect of these different components is to alter the shape of the \mbox{H\,{\sc i}} line, which can be seen clearly in \autoref{figure:PKS1830-211_hi_z089_comparison}. A net positive correlation between the $\mathrm{EW}$ of the whole \mbox{H\,{\sc i}} line and $S_{\rm c}$ indicates that the images of the core have a higher covering factor than the Einstein ring, perhaps due to the relative position of the inner \mbox{H\,{\sc i}}-depleted disc and source (see e.g. \citealt{Koopmans:2005}) or the extended nature of the ring.

In summary, we find that the EWs of the total, NE and SW  components of the \mbox{H\,{\sc i}} line fluctuate at the level of a few per\,cent over yearly time-scales and are correlated with larger fractional variations in the total continuum flux density of the source. Over the same time-scale, there is no significant evidence of variability in the ratio of the NE and the SW components of the \mbox{H\,{\sc i}} line. This behaviour naturally arises if intrinsic variability in the core causes simultaneous changes in the NE and the SW images (on time-scales larger than the known 25\,d delay) and leads to differential changes in the source surface brightness distribution with respect to the absorbing \mbox{H\,{\sc i}} gas. We propose that similar intervening \mbox{H\,{\sc i}} lines which have two or more distinct and widely spaced ($\Delta{v} \sim 100$\,km\,s$^{-1}$) velocity components, and which exhibit correlated variability, should provide a reasonably strong prior for finding new gravitationally lensed quasar candidates.

\subsubsection{A silhouette against intrinsic variability in the blazar jet}

PKS\,B1830$-$211 is known to exhibit intrinsic high-energy variability which is
typical of a luminous blazar, including several recently detected strong
$\gamma$-ray flares (\citealt{Barnacka:2011, Donnaruma:2011, Abdo:2015, Barnacka:2015}). Observations at mm wavelengths (e.g. \citealt{Garrett:1997, Jin:2003,
  Muller:2008}) also reveal intrinsic variation in the flux densities of the
NE and the SW images of the core which is consistent with the formation of new
synchrotron-emitting plasmons which are ejected at the base of
the quasar jet. The ratio of these components is then modulated by a 25\,d time delay at the gravitational
lens (\citealt{Lovell:1998, Wiklind:2001, Barnacka:2011,
  Barnacka:2015}). Evidence for intrinsic changes in the quasar core is 
  further supported by mm-imaging of rapid (weekly
and monthly) $\sim$100\,$\mu$as-scale changes in the centroid
separation and structure of these components (\citealt{Garrett:1997,
  Garrett:1998, Jin:2003}). \cite{Nair:2005} proposed a helical model for the jet to explain this behaviour, with an observed doppler-boosted precession period of approximately 1\,yr, which they suggested could be evidence for a binary supermassive black hole. More
recent observations over a few months were carried out by
\cite{Marti-Vidal:2013} using the Atacama Large Millimetre Array
(ALMA) and spanning frequencies between 100 and 300\,GHz, revealing
sub-mm flaring activity which was concurrent with a particularly
energetic $\gamma$-ray flare. Their work uncovered chromatic structure
close to the base of the jet (a core-shift effect) for which they
found that a concave-jet model of plasmon ejection could explain the
observed frequency-dependent variation in the flux-density ratio
between the two core images. 

The simultaneous brightening and fading of the \mbox{H\,{\sc i}} absorption which we find in the NE and the SW components, correlated with variability in the total continuum flux density, is a direct consequence of the cumulative formation and eventual dissipation of optically thick plasmons at the base of the jet. Brightening of the core leads to an overall concentration of the total continuum flux density towards the NE and the SW components, which then increases the EW we measure at these velocities at the expense of the non-variable steep-spectrum ring. We expect that further 21-cm line observations with ASKAP, at higher spectral S/N and cadences, would reveal the effect of the 25\,d lens time delay in the NE/SW EW ratio.

\subsubsection{Comparison with the molecular line variability}

Similar long-term variability has also been observed in
molecular absorption at mm-wavelengths. \cite{Muller:2008} carried out
monitoring between 1995 and 2007 of the
\mbox{HCO$^{+}$\,(2-1)} and \mbox{HCN\,(2-1)} lines at 3-mm using the
IRAM Plateau de Bure interferometer (PdBI) and IRAM 30\,m
telescopes. During this period, they found significant, and correlated,
variability in molecular absorption towards the NE and the SW
lens components and the total continuum flux density. As with the
\mbox{H\,{\sc i}} absorption, this behaviour is symptomatic of changes in the source brightness distribution. However, in contrast to the \mbox{H\,{\sc i}} data presented
here, they also found significant changes in the opacity ratio of the NE component and the blue-wing of the SW component, on time-scales much larger than that expected from the known 25\,d lens time delay. Notably, they found a decrease of almost 50\,per\,cent 
in the NE/SW ratio between 2003 and 2006, the direct result of a reduction in absorption against the NE component by more than a factor of 6. Given that the sub-structure of the imaged-core is known to change on scales of more than $\sim 100\,\mu$as ($\sim 1$\,pc at $z = 0.886$) at mm-wavelengths (e.g. \citealt{Jin:2003}), over time-scales of several months to a year,
\cite{Muller:2008} proposed that this behaviour is the result of changes in the NE and the SW sight lines through a sparse distribution of diffuse molecular clouds. 

Molecular line variability has continued to be seen in subsequent observations of both the NE and the SW source components. \cite{Muller:2013} compared their ATCA 7-mm spectra from earlier observations in 2009 and 2010 obtained by \cite{Muller:2011}, and found significant differences 
for several species. \cite{Schulz:2015} used the Effelsberg 100-m telescope at cm-wavelengths to monitor variability in \mbox{CS\,(1-0)} absorption towards the SW component, finding a striking similarity to the   variations in \mbox{HCO$^{+}$\,(2-1)} discovered by \cite{Muller:2008}. Recently, \cite{Muller:2014} carried out a spectroscopic survey of common interstellar molecules using the the ALMA. They detected short-term ($\sim$2\,months) variations in the wings of the saturated
\mbox{CO\,(4-3)}, \mbox{HCO$^{+}$}, and \mbox{HCN\,(2-1)} lines towards
the SW image which were concurrent with the sub-mm counterpart to the
$\gamma$-ray flare found by \cite{Marti-Vidal:2013}. They also
compared their ALMA spectra with the previous ATCA 3 and 7-mm spectra
(obtained by \citealt{Muller:2008, Muller:2011, Muller:2013}), seeing again the approximately
yearly variability in several other molecular species in addition to that in \mbox{HCO$^{+}$}, \mbox{HCN}, and \mbox{CS}. However, it remains to be seen if the spectral variability seen in either the \mbox{H\,{\sc i}} or molecular lines is periodic, as might be expected from the helical jet model proposed by \cite{Jin:2003}.

Relative changes in the NE and the SW opacities of the molecular lines are consistent with optical depth variations which are well matched to $\sim$1\,pc-scale changes in the source structure at mm-wavelengths. The absence of similar variability in the \mbox{H\,{\sc i}} NE/SW ratio implies that structural changes in the source at 21-cm wavelengths are not sensitive to fluctuations in the optical depth. This could be due to be the typical size on which coherent opaque \mbox{H\,{\sc i}} structures are seen through the face-on disc of this galaxy, which would imply a transverse physical scale larger than a parsec. This would be consistent with high spatial-resolution imaging of several redshifted 21-cm absorbers, which typically imply sizes for coherent absorbing structures ranging from the 10\,pc to kpc-scale (see e.g. \citealt{Lane:2000, Srianand:2013, Borthakur:2014,  Biggs:2016, Dutta:2016}). A detailed 21-cm study of the Local Group galaxies M31, M33 and LMC by \cite{Braun:2012} revealed self-opaque high-$N_{\rm HI}$ structures with typical sizes $\sim$100\,pc. \cite{Gupta:2012} similarly concluded from their 21-cm survey of a large sample of \mbox{Mg\,{\sc ii}} absorbers towards compact quasars that the typical correlation length $\sim$30-100\,pc. We do not find evidence for smaller parsec-scale structures in the absorbing gas, although this may simply be an issue of resolution. Variable components of the source at 21-cm may be significantly more extended than at mm-wavelengths, effectively not resolving small-scale variations in the \mbox{H\,{\sc i}} opacity. Unfortunately, the source is significantly scatter broadened by the Galactic ISM at cm-wavelengths (\citealt{Jones:1996}) and so it is difficult to directly measure the size seen by the absorber. Assuming that the core is emitting in the optically thick regime, we scale the $\sim$100\,$\mu$as measured by \cite{Jin:2003} at 43\,GHz by $r \propto \nu^{-0.83}$ to estimate a typical scale $\sim$3\,mas ($\sim$20\,pc at $z = 0.886$) for changes in the magnified core at 753\,MHz. We do, however, caution that the time sampling and sensitivity of the current \mbox{H\,{\sc i}} data are limited and our understanding of this system would be significantly improved by additional frequent monitoring of this source over the next five years with either ASKAP or a radio telescope of similar capability.

\subsubsection{Limits on the absorbing structure from ISS}\label{section:scintillation}

In addition to the approximately yearly variability driven by changes
in the flux density of the background source, the presence of
sufficiently small-scale ($\lesssim 1\,$mas) structure in the
absorbing medium could drive stochastic fluctuations in the \mbox{H\,{\sc i}} EW on time-scales of several months (and even years), due to
scintillations associated with the turbulent ISM of
the Milky Way (\citealt{Macquart:2005}). This variability is caused by
the response of ISS to the presence of
small-scale structure imprinted on the image of the absorbed source,
and may therefore be evident even if the background source is not
sufficiently compact to scintillate itself. The presence (or absence)
of variability due to ISS can
therefore be used to constrain the sizes and distribution of absorbing
\mbox{H\,{\sc i}} structures in the intervening galaxy. In the direction of PKS\,B1830$-$211, the NE2001 scattering model of the Galactic ISM by
\cite{Cordes:2002} predicts that the angular scale probed by
refractive ISS is $\theta_{\rm ref} = 2.1\,$mas
at the observed \mbox{H\,{\sc i}} frequency of 753\,MHz. The
corresponding predicted point-source modulation index (i.e.~the rms of
the variations normalized by the mean flux density), due to refractive
scintillation, is $m_{\rm pt}=12\%$ and the predicted time-scale is $710\,v_{30}\,$\,d, where $v_{30}$ is the speed of the ISM
transverse to the line of sight normalized to
30\,km\,s$^{-1}$.

At the redshift of the absorbing system, refractive scintillation is
sensitive to absorbing \mbox{H\,{\sc i}} structure on linear
scales less than approximately $16$\,pc. If a single absorbing cloud
of size $\theta_{\rm S}$ were present across the extent of the background source, the optical depth of the absorption line would
exhibit apparent variations with a modulation index of $m_{\rm pt}$ if
$\theta_{\rm S} < \theta_{\rm ref}$ or $m_{\rm pt} (\theta_{\rm ref}/\theta_{\rm S})$ otherwise.  However, if the
background source is large relative to the typical cloud size, then it
is more likely that $N$ absorbing clouds are present across the extent
of the source, and the modulation index would be reduced by a factor
of $\sim N^{1/2}$ as the individual variations due to each absorbing
system add incoherently with those of other clouds.

In \autoref{figure:variability_probability}, we showed the probability of residual stochastic variability in the EW data, once the yearly time-scale continuum variability has been taken into account. Assuming our model for continuum-driven variability, we obtain a 99.7\,per\,cent (3\,$\sigma$) upper limit of approximately 6\,per\,cent on any residual variability due to ISS. This limit implies that the characteristic
scale of the \mbox{H\,{\sc i}} absorbing structure exceeds
approximately $38\,$pc across each lensed component, consistent with the
conclusions reached in the previous subsection. Another possible
scenario is that the scintillations are instead quenched by the
contributions of many absorbing structures across the angular extent
of the lensed image. In that case, refractive scintillations
associated with any one absorbing cloud are diluted by the independent
variations associated with other clouds. The present limit on optical depth variations therefore implies that six or more compact structures of size $\lesssim 16\,$pc need to be present to reduce the modulation index
of the optical depth variations below a detectable level. The foregoing limits apply if the range of velocities spanned by the absorbing clouds is comparable to the EW, $v_{\rm EW}$, or if all the clouds which contribute to the observed absorption are contained within an angular scale $\theta_{\rm ref}$.

If the absorbing clouds span a velocity range small compared to the EW then the variations in the individual spectral lines which comprise the ensemble absorption spectrum would fluctuate independently, and so reduce the overall amplitude of the variations in the EW provided the clouds are distributed across a region exceeding $\theta_{\rm ref}$, and are correlated neither in velocity nor space. Denoting $\Delta v$ as the velocity range over which scintillations are completely independent between sets of absorbing clouds, the amplitude of the variations is reduced by a factor $\sim \sqrt(v_{\rm EW}/\Delta v)$, since their contributions would sum together incoherently.  In other words, the contribution of several clouds at different velocities reduces the same variability in the same manner that an extended spatial distribution of clouds does.  To this end, the lower limit on the number of compact clouds contributing to the opacity variations remains the same.

We note further that any stochastic variability due to scintillation would also imply variations in the NE/SW EW ratio, and the limit on the variation of this ratio places a similar limit on the contribution of scintillations to the variability. 

\section{MG\,J0414+0534}\label{section:mgj0414+0534}

MG\,J0414$+$0534 is a heavily dust-reddened quasar at $z = 2.639 \pm 0.002$ (\citealt{Lawrence:1995a}), which is gravitationally
lensed into four image components (\citealt{Hewitt:1992}) by a foreground early-type galaxy at $0.9584 \pm 0.0002$ (\citealt{Tonry:1999})
and possibly a second `Object X' at unknown redshift
(\citealt{Schechter:1993,Angonin-Williame:1994,Falco:1997}). Notably
this quasar hosts the most distant and luminous water maser ever
discovered, consistent with an extremely dusty and gas-rich host
galaxy (\citealt{Impellizzeri:2008}), which is further corroborated by the presence of \mbox{H\,{\sc i}} absorption (\citealt{Moore:1999}). The mid-infrared and radio
flux ratios between the brightest components of the image are found to be anomalous when considering just a simple model of the lens, consisting of only the
early-type galaxy and Object X components. This implies that several
undiscovered components for the lens may exist, either in the form of satellite
sub-haloes at the redshift of the early-type galaxy
(e.g. \citealt{Minezaki:2009, MacLeod:2013}) or as intervening
line-of-sight sub-structure (e.g. \citealt{Inoue:2012,
  Inoue:2015}). Therefore, detections of intervening galaxies through line-of-sight absorbing 
  \mbox{H\,{\sc i}} gas may provide useful
constraints for selecting between these models.

\begin{figure}
\centering
\includegraphics[width=0.45\textwidth]{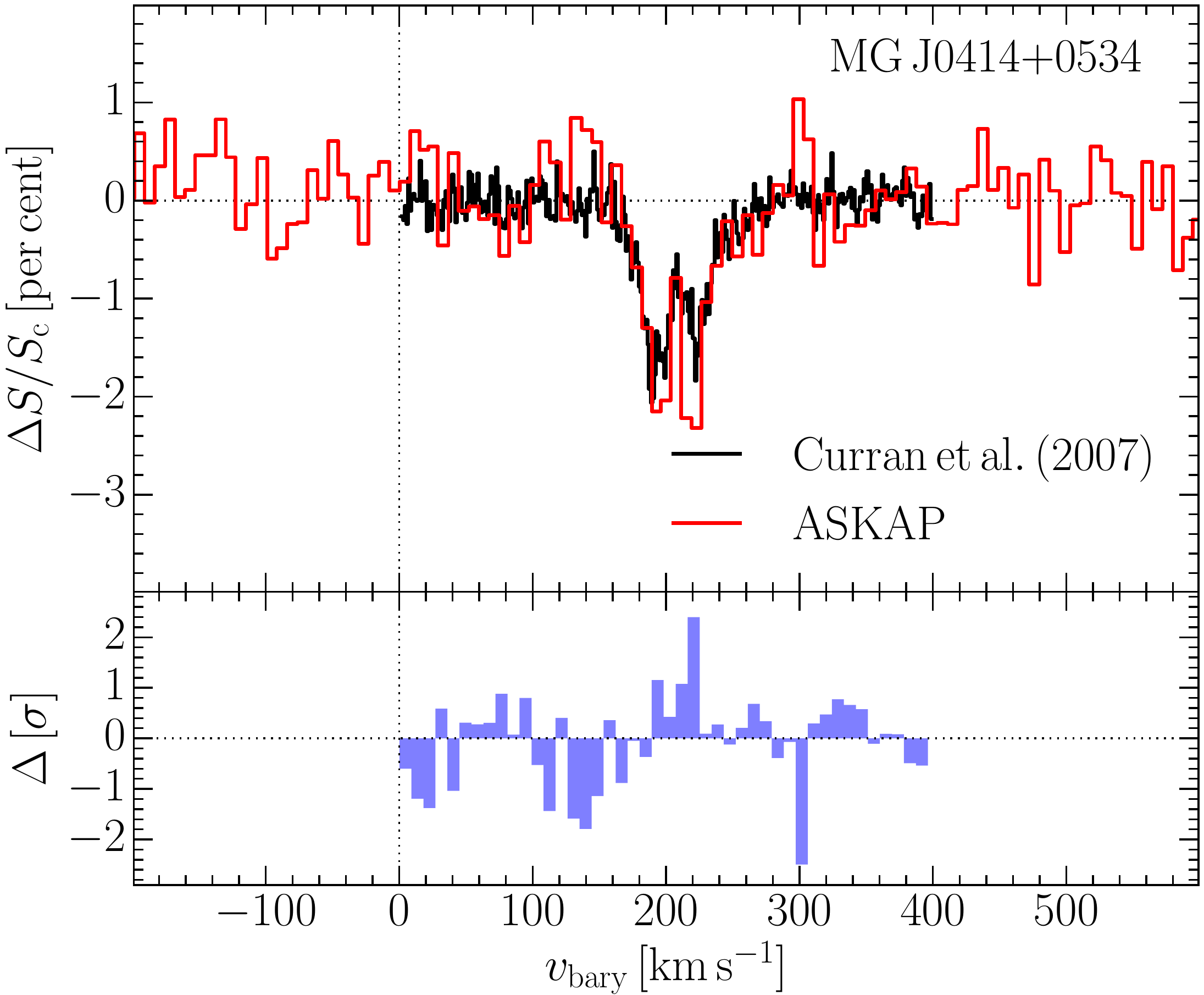}
\caption{Spectra from ASKAP BETA (this work) and GBT (\citealt{Curran:2007b}) showing \mbox{H\,{\sc i}} absorption associated with the
  early-type lensing galaxy at $z = 0.9584$ towards
  MG\,J0414$+$0534. The difference spectrum (blue filled)
  is shown in the bottom
  panel.}\label{figure:MGJ0414+0534_hi_z096_comparison}
\end{figure}

Using the Green Bank Telescope (GBT), \cite{Curran:2007b} discovered
\mbox{H\,{\sc i}} absorption at the redshift of the early-type lensing
galaxy. They speculated that the anomalous $\sim$200\,km\,s$^{-1}$ velocity
offset with respect to system redshift of $z = 0.9584 \pm 0.0002$ may arise from gas which is accreting on to the galaxy. Alternatively, the gas may be associated with a
separate satellite galaxy at the same redshift, which has not yet been
identified in either the optical or the IR data. We also detected this line with  
ASKAP BETA, and find the spectrum to be consistent (albeit at much lower
spectral resolution) with the velocity components originally
identified by Curran et al. (see
\autoref{figure:MGJ0414+0534_hi_z096_comparison}). However, we did not
detect any of the other \mbox{H\,{\sc i}} absorbers previously reported in the literature. 
We find that the system at $z = 0.379$, reported by \cite{Curran:2011c} as a possible
counterpart to the putative Object X, is completely obscured by 1030\,MHz RFI caused by the secondary surveillance radar uplink frequency used
in air traffic control services worldwide. Given the ubiquity of this
RFI signal at other observatory sites, we suspect that the original
detection using the GBT may have been confused with a
negative residual feature generated from RFI during the standard practice of single-dish on-off switching. We also did not detect the 21-cm absorbers reported by \cite{Tanna:2013} at $z = 0.339$
and $z = 0.534$, which are expected to have peak optical depths of 0.6 and 0.45,
respectively. We do not see any detectable RFI in our data at these frequencies\footnote{This could simply be the result of differences in the radio frequency environments of the two observatories.} and the optical depth sensitivity ($\sigma_{\rm \tau} \sim 0.005$ in a single 18.5\,kHz channel) is sufficient for detection. Based on our results we conclude that there is no evidence for further intervening \mbox{H\,{\sc i}} absorbers at $z \leq 1$, beyond that associated with the known lensing galaxy at $z = 0.958$.

\begin{figure}
\centering
\includegraphics[width=0.45\textwidth]{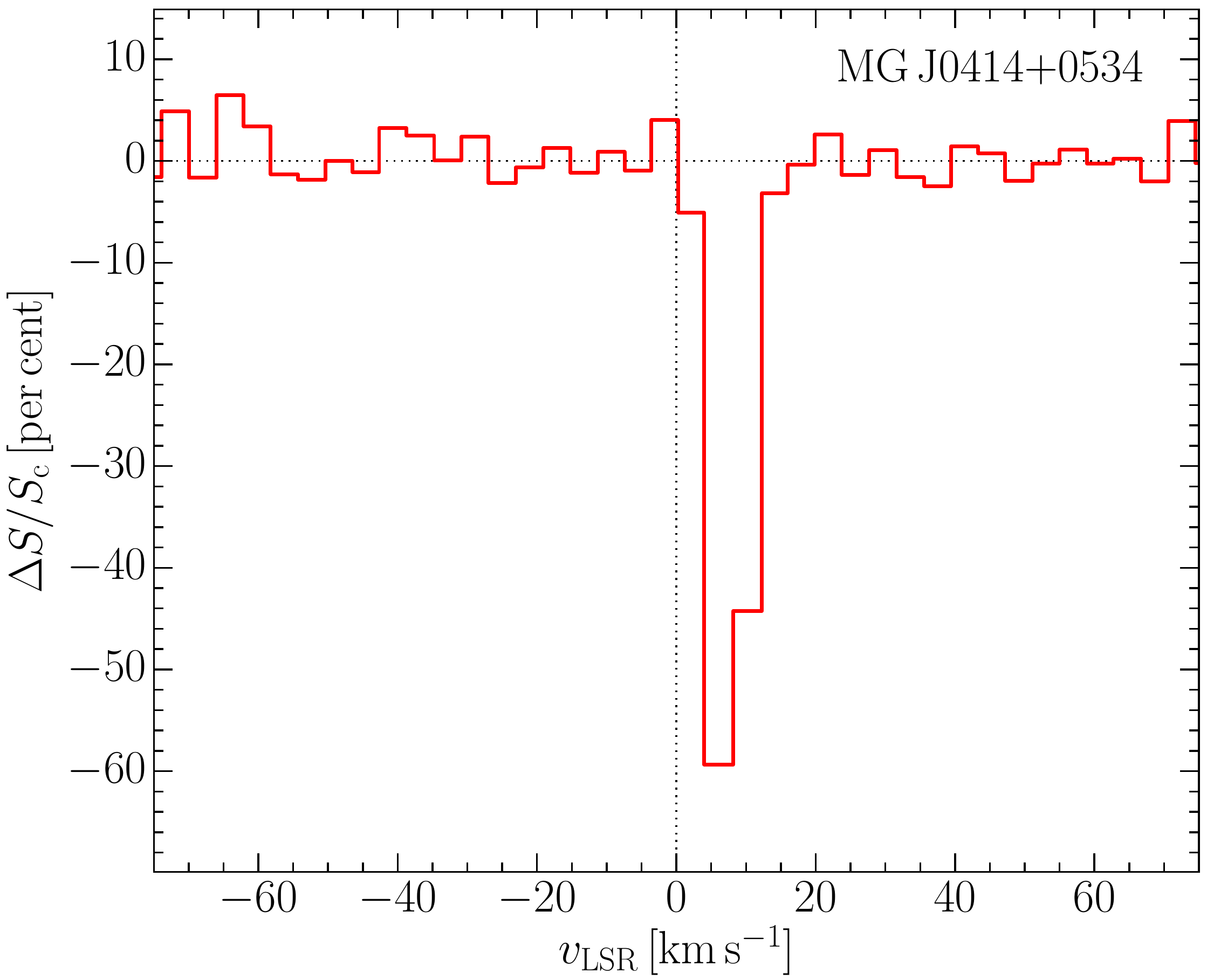}
\caption{ASKAP BETA spectrum showing Galactic \mbox{H\,{\sc i}}
  absorption towards
  MG\,J0414$+$0534.}\label{figure:MGJ0414+0534_hi_mw}
\end{figure}

\begin{figure*}
\centering
\includegraphics[width=1.0\textwidth]{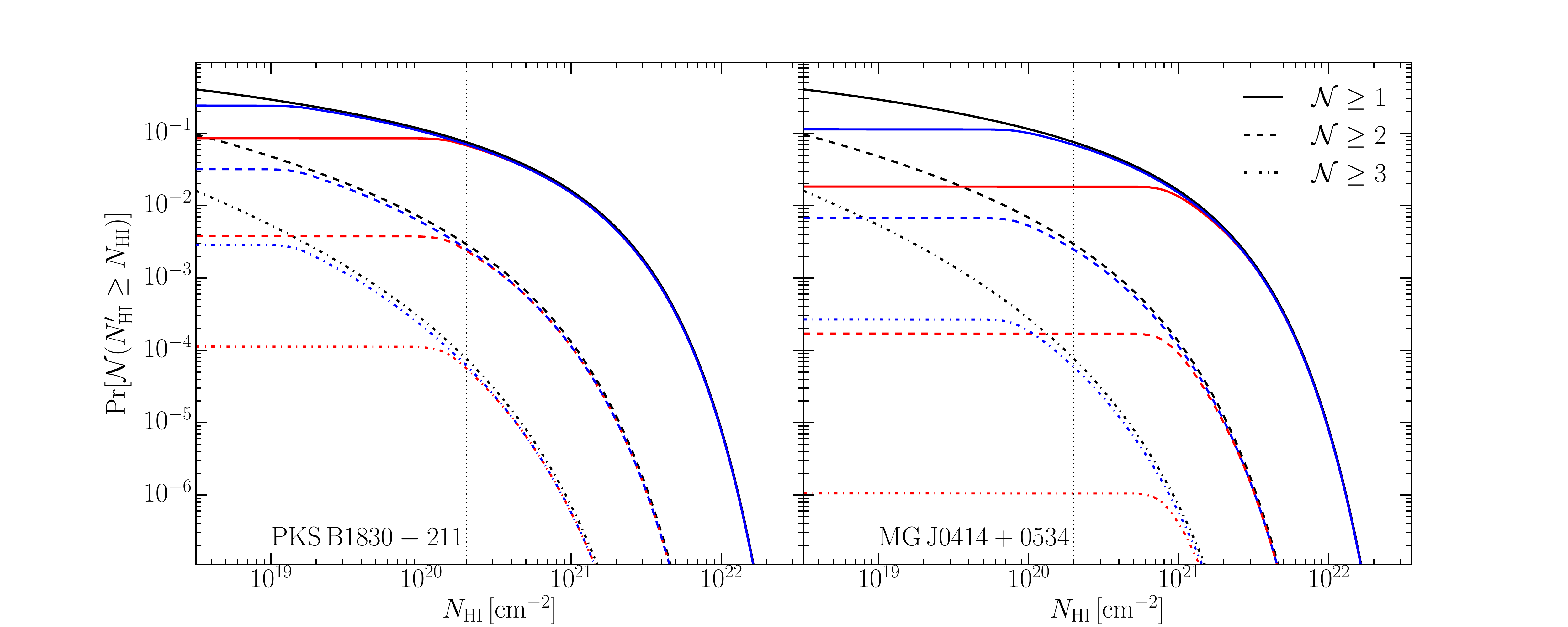}
\caption{The probability of $\mathcal{N}$ or more intervening
  \mbox{H\,{\sc i}} systems occurring along a single random sight-line
  between $z = 0$ and $1$, where the black lines denote the cumulative
  distribution as a function of \mbox{H\,{\sc i}} column density based on 21-cm emission and DLA surveys. The blue and red lines denote the probability of detection in our
  averaged ASKAP BETA data for spin temperatures of $100$ and
  $1000$\,K, respectively. We use a simple Gaussian velocity model for the
  absorbing \mbox{H\,{\sc i}} gas, assuming a line FWHM of
  $50$\,km\,s$^{-1}$ and a covering factor of unity. The vertical
  dotted line represents the minimum column density which defines a
  DLA.}\label{figure:absorber_probability}
\end{figure*}

Apart from clarifying the extragalactic sight line towards MG\,J0414$+$053,
we also detected \mbox{H\,{\sc i}} absorption associated with the Milky Way. In
\autoref{figure:MGJ0414+0534_hi_mw}, we show a section of the
ASKAP BETA spectrum centred on the LSR frame. At $b =
-31\degr$ and $l = 187\degr$, the quasar line-of-sight is positioned
somewhat out of the plane and towards the Galactic anti-centre. The
LSR velocity of the peak absorption ($\approx +5$\,km\,s$^{-1}$) is consistent with this direction through the Galactic plane. The narrow line width, and lack of significant substructure compared with the PKS\,B1830$-$211 line, is also consistent with a shorter path length through the \mbox{H\,{\sc i}} disc of the Galaxy.

\section{The expected frequency of intervening
  absorbers}\label{section:expected_number}

A primary goal of future broad-band wide-field 21-cm surveys for intervening absorption will be to measure the cosmological
\mbox{H\,{\sc i}} mass density between $z \sim 0.2$ and $1.7$,
enabling direct comparison of the evolution of atomic gas in galaxies
with that of the global star formation rate and molecular mass density
(e.g. \citealt{Kanekar:2004, Morganti:2015}). Although the mass density of atomic gas
in DLAs has been measured at these redshifts using the \emph{HST}, 
the sample sizes are small and possibly subject to selection bias
(\citealt{Rao:2006, Neeleman:2016}). The first attempt to demonstrate
the feasibility of measuring the atomic mass density with a genuinely radio-selected 21-cm
absorption survey was carried out by \cite{Darling:2011} (see also \citealt{Darling:2004}). Using pilot
data from the Arecibo Legacy Fast Arecibo L-band Feed Array (ALFALFA)
survey, Darling et al. were able to obtain an upper limit to the
\mbox{H\,{\sc i}} column density frequency distribution at low
redshift ($z \lesssim 0.06$) which was consistent with the 
distribution measured by \cite{Zwaan:2005} from 21-cm emission in a sample of nearby galaxies.

The substantial redshift interval covered by the ASKAP BETA data towards PKS\,B1830$-$211 and MG\,J0414$+$0534 allows us to carry out a test of whether the number of detected intervening absorbers
is consistent with current measurements of the column density
frequency distributions at low and high redshift. We follow the method
of A15 to calculate the expected number of
intervening systems of a given \mbox{H\,{\sc i}} column density. Over the redshift range covered by our data we interpolate between the analytical $\Gamma$-functions for the column density distribution measured from 21-cm emission in the nearby Universe (\citealt{Zwaan:2005}) and
DLAs at $z \sim 3$ (\citealt{Noterdaeme:2009}), extrapolating to lower column densities. Using the expected number of detections we calculate the Poisson probability of $\mathcal{N}$ or more intervening
\mbox{H\,{\sc i}} systems intercepting a single random sight-line at
redshifts between 0 and 1. We then adopt a simple Gaussian velocity
model for the absorbing \mbox{H\,{\sc i}} gas to calculate the column density sensitivity of our ASKAP BETA spectra (as shown in \autoref{figure:PKS1830-211_beta_spectrum} and \autoref{figure:MGJ0414+0534_beta_spectrum}) and hence estimate the
probability of detecting $\mathcal{N}$ or more  absorbers towards
PKS\,B1830$-$211 and MG\,J0414$+$0534. In our model, we assume an FWHM line width
equal to 50\,km\,s$^{-1}$ (equal to the typical widths seen in the
detections towards these quasars), a covering factor of 1, and a spin
temperature in the typical range of 100 - 1000\,K. 

The resulting probability distributions, as a cumulative function of column density, are shown in \autoref{figure:absorber_probability}. If our sight lines were randomly chosen, we could simply test our detection yields against these estimated probabilities. However, given that we selected our sources based on the knowledge that they are strongly gravitationally lensed by at least one intervening gas-rich galaxy, within the redshift range covered by our data, we should ignore the primary lensing galaxy in our analysis. With this in mind, the detection yield in the case of PKS\,B1830$-$211 is one intervening absorber, while for MG\,J0414$+$0534 it is zero. Both cases, while not hugely constraining, are entirely consistent with that expected from our current understanding of the \mbox{H\,{\sc i}} content of the Universe. We estimate that the probability of detecting two or more absorbers towards PKS\,B1830$-$211 is less than about 3\,per\,cent, which seems consistent with the absence of any new detections towards this system. With an estimated probability of less than 0.03\,per\,cent, we agree with \cite{Tanna:2013} that the probability of detecting three or more intervening absorbers towards MG\,J0414$+$0534 (in addition to the lensing galaxy) is very unlikely. Our falsification of these original detections is therefore consistent with expectations.

\section{Conclusions}\label{section:conclusions}

We have carried out a wide-band 21-cm study of the lensed radio
quasars PKS\,B1830$-$211 and MG\,J0414$+$0534 using commissioning data
from the Boolardy Engineering Test Array of the Australian SKA
Pathfinder. Our data allowed us to survey these sight-lines for cold
gas revealed by \mbox{H\,{\sc i}} and OH absorption over an almost
continuous redshift interval between $z_{\rm HI} = 0$ and 1. We summarize our 
conclusions as follows:
\begin{enumerate}
\item\noindent Towards PKS\,B1830$-$211 we have re-detected \mbox{H\,{\sc i}} and OH absorption
  associated with the $z = 0.886$ lens (\citealt{Wiklind:1996, Chengalur:1999}) and the second absorber at $z = 0.193$ (\citealt{Lovell:1996}). No further intervening galaxies were detected 
  in this sight line over these redshifts.
\item \noindent In the case of MG\,J0414$+$0534, we did not re-detect
  three of the intervening systems previously reported in the
  literature (\citealt{Curran:2011c, Tanna:2013}), confirming only the
  \mbox{H\,{\sc i}} absorber associated with the early-type
  gravitational lens at $z = 0.958$ \citep{Curran:2007b}. Given that
  our data are sufficiently sensitive to detect these lines, it is likely that the original identifications were mistakenly
  confused with RFI. We conclude that no
  \mbox{H\,{\sc i}} counterparts have yet been found for the intervening 
  sub-structures which could explain the observed flux ratio anomalies between image components.
\item\noindent In the $z = 0.886$ lensing galaxy of PKS\,B1830$-$211,
  we find evidence of approximately yearly variability of a few per\,cent over the timeline spanned by WSRT  observations
  undertaken in 1997/1998 and our ASKAP BETA observations in
  2014/2015. Variations in the \mbox{H\,{\sc i}} EW are correlated
  with the total continuum flux density and appear to be coherent between the NE and the SW images of the core. This behaviour arises naturally as a result of intrinsic variability in the compact core which is preferentially covered by the absorbing gas. We suggest that similar behaviour in absorption lines with several widely spaced velocity components detected in the 21-cm spectra of quasars could be used as a strong prior for selecting gravitationally lensed candidates.
\item\noindent The absence of any detectable variability in the ratio of
  \mbox{H\,{\sc i}} EWs towards the NE and the SW
  components, in contrast to that seen in the molecular absorption
  lines, suggests that the absorbing gas is distributed on scales
  larger than that probed by changes in the source structure. Higher cadence and S/N observations over
  the next five years with ASKAP will be able to test for further
  changes in optical depth and provide more stringent constraints on this
  model and the physical size of the \mbox{H\,{\sc i}} gas
  distribution.
\item\noindent The absence of any detectable stochastic \mbox{H\,{\sc
      i}} variability from ISS lends further evidence that opaque \mbox{H\,{\sc i}} structures in the foreground galaxy are distributed on physical scales larger than approximately 40\,pc.
\item\noindent Given the 21-cm optical depth sensitivity of our data,
  and assuming a simple Gaussian velocity model for the absorbing
  \mbox{H\,{\sc i}} gas, we find that the number of intervening
  systems detected towards both PKS\,B1830$-$211 and MG\,J0414$+$0534 is consistent with the column
  density frequency distribution functions measured from local 21-cm
  and high-$z$ DLA surveys.
\end{enumerate}

\section*{Acknowledgements} 

JRA thanks Ron Ekers, Paolo Serra, Ger de Bruyn and Naomi
McClure-Griffiths for useful discussions on interpretation, Ryan Shannon and David Parkinson for advice and comments on the
statistical analysis, and Jim Lovell for providing digital versions of his published spectra. We also thank Sebastien Muller and the anonymous referee for comments which helped improve this paper. 

The Australian SKA Pathfinder is part of the Australia Telescope
National Facility which is managed by CSIRO. Operation of ASKAP is
funded by the Australian Government with support from the National
Collaborative Research Infrastructure Strategy. Establishment of the
Murchison Radio-astronomy Observatory was funded by the Australian
Government and the Government of Western Australia. ASKAP uses
advanced supercomputing resources at the Pawsey Supercomputing
Centre. We acknowledge the Wajarri Yamatji people as the traditional
owners of the Observatory site.

JRA acknowledges support from a Bolton Fellowship. Parts of this
research were conducted by the Australian Research Council Centre of
Excellence for All-sky Astrophysics (CAASTRO), through project number
CE110001020. We have made use of \texttt{ASTROPY}, a
community-developed core \texttt{PYTHON} package for astronomy
(\citealt{Astropy:2013}); the NASA/IPAC Extragalactic Database (NED)
which is operated by the Jet Propulsion Laboratory, California
Institute of Technology, under contract with the National Aeronautics
and Space Administration; NASA's Astrophysics Data System
Bibliographic Services; and the VizieR catalogue access tool operated
at CDS, Strasbourg, France.




\bibliographystyle{mnras}
\bibliography{/Users/jra/reports_thesis_papers/bibliography/james}

\bsp	
\label{lastpage}
\end{document}